\documentclass[aps,pra,superscriptaddress,twocolumn]{revtex4-1}

\usepackage{graphicx}
\usepackage{epstopdf}

\begin{document}

\title{Photon blockade with a four-level quantum emitter coupled to a photonic-crystal nanocavity}

\author{M.~Bajcsy}
\email{bajcsy@stanford.edu}

\author{A.~Majumdar}

\author{A.~Rundquist}

\author{J.~Vu\v{c}kovi\'{c}}
\affiliation{Ginzton Laboratory, Stanford University, Stanford, CA 94305}

\date{\today}

\begin{abstract}
We study the photon blockade phenomenon in a nanocavity containing a single four-level quantum emitter. 
By numerically simulating the second-order autocorrelation function of the intra-cavity field with realistic parameters achievable in a state-of-the-art photonic-crystal nanocavity, we show that significant photon blockade effects appear even outside of strong coupling regime. We introduce an intuitive picture of the photon blockade that explains the performance difference between the two-level and the four-level emitter schemes reported in previous works, as well as why -- in contrast to a cavity containing a two-level atom -- signatures of photon blockade appear and should be experimentally observable outside the strong coupling regime when a four-level emitter is used. Finally, we show that thanks to the emitter-cavity coupling achievable in a nanocavity, photon blockade can be realized despite the large frequency difference between the relevant optical transitions in realistic four-level emitters,  which has so far prevented experimental realization of this photon blockade scheme.

\end{abstract}

\maketitle

\section{Introduction}

Photon blockade, in which the transmission of only one photon through a system is possible while excess photons are absorbed or reflected, is a concept first proposed by Imamo\u{g}lu \textit{et al.} \cite{Imamoglu1997} in analogy with the Coulomb blockade effect of electron transport through mesoscopic devices \cite {Fulton1987}. 
In recent years, it has generated a lot of  scientific interest as a promising tool for controllably implementing repulsive interactions between photons, which can be used for quantum simulations \cite{Greentree2006, Hartmann2006, Carusotto2012}.  Additionally, photon blockade provides a mechanism for coherent generation of non-classical light states \cite{Faraon2008a, Majumdar2012a} and for photon routing \cite{Rosenblum2011} for applications such as quantum information processing and quantum cryptography \cite{Vuckovic2009}.
Numerous platforms have been proposed for experimental implementation of photon blockade, such as single quantum emitters coupled to cavity quantum electrodynamic (cQED) systems \cite{Carmichael1992, Imamoglu1997, Kimble2005, Kimble2008, Majumdar2012, Liew2010} or nanowire-guided surface plasmons \cite{Chang2007}, atoms in Rydberg states \cite{Gorshkov2011, Dudin2012, Peyronel2012}, and atomic ensembles in photonic waveguides \cite{Chang2008}. Besides the recent experiments involving Rydberg atoms \cite{Dudin2012, Peyronel2012}, experimental demonstrations of photon blockade have so far been achieved for the most part with cQED platforms in the regime of strong coupling. These reports include atomic cQED systems based on a Fabry-Perot cavity \cite{Kimble2005} and solid-state cQED systems based on single quantum dots embedded in a photonic crystal nanocavity \cite{Faraon2008a, Reinhard2012}.  Additionally, cQED systems achieving photon blockade in the weak coupling regime have been demonstrated with laser cooled atoms coupled to a toroidal microcavity \cite{Kimble2008} and predicted for quantum dots coupled to either a bimodal nano-cavity \cite{Majumdar2012} or to a photonic molecule \cite{Liew2010, Bamba2011}. Unfortunately, platforms implementing photon blockade have so far proved to be difficult to scale into more complex systems. Photonic crystal cavities containing strongly-coupled self-assembled quantum dots have shown a lot of promise in this area \cite{Faraon2011}, but further scalability of this platform still remains challenging due to the non-deterministic spatial location and spectral properties of the self-assembled quantum dots. 
\begin{figure}
   \begin{center}
  \begin{tabular}{c}
     \includegraphics[width=7cm]{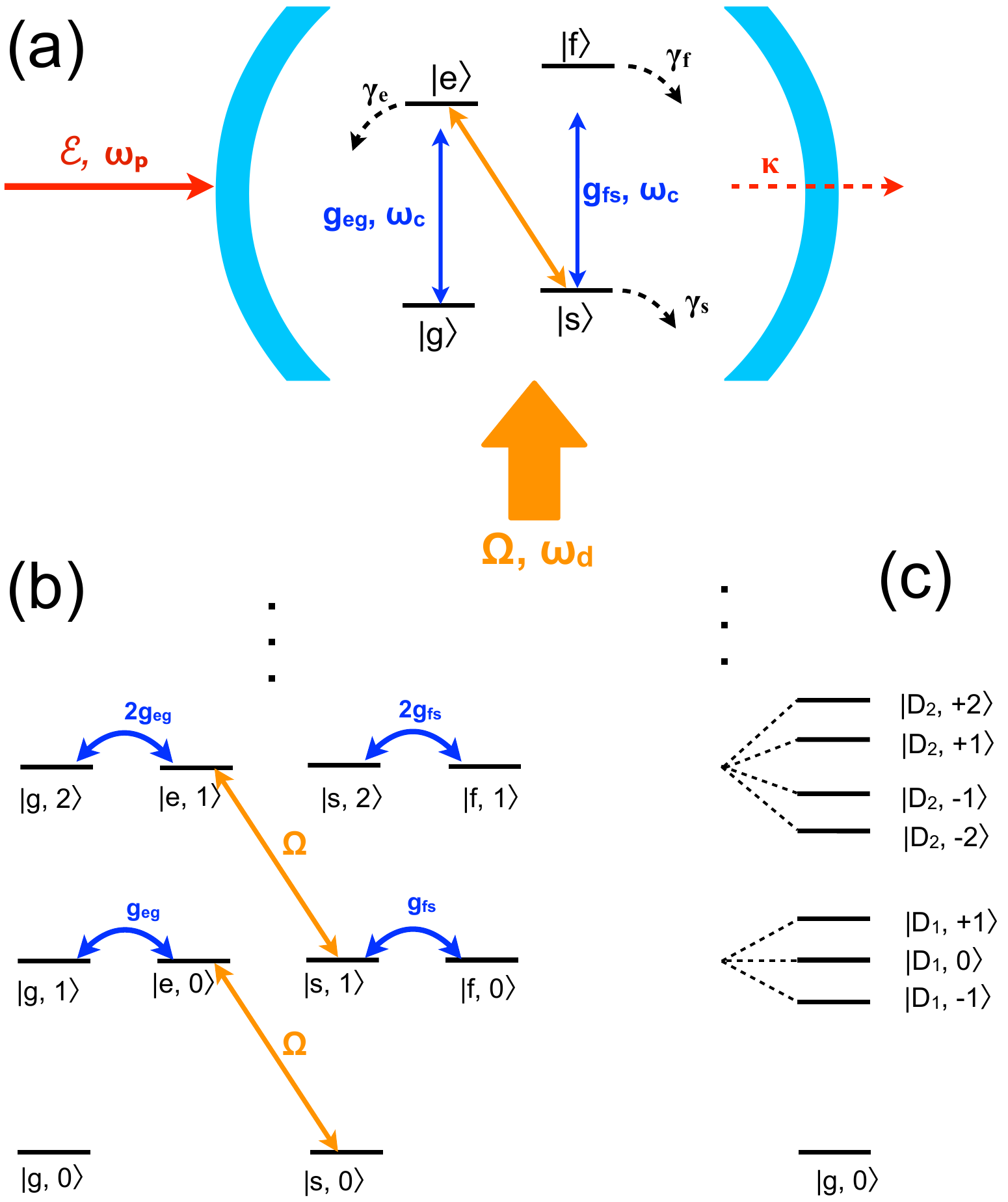}
   \end{tabular}
   \end{center}
   \caption{\label{fig1} (Color online)  (a) Schematics of the described  photon blockade system consisting of a four level quantum emitter coupled to a cavity. The cavity is driven by a weak classical probe with frequency $\omega_p$ and driving rate $\mathcal{E}$, , while the transitions from $\vert g\rangle$ to $\vert e\rangle$ and from $\vert s\rangle$ to $\vert f\rangle$ are coupled to the cavity field. Levels $\vert s \rangle$ and $\vert e\rangle$ are coupled by a classical field (drive) that does not couple into the cavity and has a frequency $\omega_d$ and driving rate $\Omega$.  (b) The bare states of the overall system with quantized cavity field and the coupling rates between them. (c) The ground state, first, and second manifold of the dressed states of the overall system for the idealized atom with $\Delta_c=\Delta_{sg}=\Delta_{eg}=\Delta_{fg}=0$ (i.e., probe resonant with the cavity, drive resonant with the $\vert e\rangle \rightarrow \vert s\rangle$ transition, and cavity resonant with $\vert g\rangle \rightarrow \vert e\rangle$ and $\vert s\rangle \rightarrow \vert f\rangle$ transitions). }
\end{figure} 
On the other hand, cold atoms are by nature identical, which makes them excellent candidates for quantum emitters in scalable architectures. However, building scalable architecture based on atomic cavity QED systems demonstrated so far is non-trivial.   Coupling laser-cooled atoms to planar photonic-crystal cavities was first suggested in Ref. \cite{Vuckovic2001} and has been since explored in additional proposals, such as \cite{Lev2004, Greentree2006}. While experimental realization of these proposals is challenging, recent developments in design of planar photonic-crystal cavities, such as those described in  \cite{Yamamoto2008, Quan2011, Li2011} have opened additional prospects of cold atoms interacting with the field maximum of a cavity with a mode volume that is ordinarily associated exclusively with solid-state cQED. 
 

In the experimental demonstrations reported so far, photon blockade was achieved with a two-level quantum emitter (a neutral atom or a quantum dot) coupled to the cavity. However, Imamo\u{g}lu \textit{et al.} originally coined the term ``photon blockade" in a proposal envisioning a four-level atom coupled to a Fabry-Perot cavity \cite{Imamoglu1997}. In the strong coupling regime, this scheme was predicted to result in a more robust photon blockade compared to the conventional approach of a two-level emitter coupled to a single mode of a cavity, with quantum interference suggested as a possible cause \cite{Werner1999}.  In the years following the initial proposal, photon blockade based on a four-level quantum emitter coupled to a cavity has been analyzed extensively  \cite{Rebic1999, Gheri1999, Werner1999, Greentree2000, Rebic2002a, Rebic2002b}. However, these theoretical studies examined the blockade under conditions observed in atomic cQED experiments in the regime of strong coupling where the rate of atom-field coupling $g$ far exceeds the cavity field decay rate $\kappa$ and, in most cases, only for idealized atoms with the transitions $\vert g\rangle \rightarrow \vert e\rangle$ and $\vert s\rangle \rightarrow\vert f\rangle$  (Fig. \ref{fig1}(a)) having the same or nearly the same frequency.

Here, we take a closer look at a four-level quantum emitter coupled to a cavity as shown in Fig. \ref{fig1}(a), in the regime of $g\approx \kappa$ with the cavity mode volume $V_{mod} \sim \lambda^3$, where $\lambda$ is the wavelength of the probe light. This regime is encountered for self-assembled quantum dots embedded in photonic-crystal microcavities, e.g.  \cite{Hennessy2007, Faraon2008a}, and can also be expected for the first generation of experiments coupling laser-cooled atoms to these cavities. Note that as an alternative to being trapped at and interacting with the field maximum of an air-mode cavity, such as those described in \cite{Yamamoto2008, Quan2011, Li2011}, the atom could also interact with the evanescent field of a dielectric-mode photonic-crystal cavity while being trapped near the surface of such cavity, e.g. using techniques described in \cite{Chang2008} or in \cite{Dawkins2011, Lacroute2012}. 

As a model system for our numerical simulation, we assume a single quantum emitter based on a laser-cooled cesium atom interacting with the field of a solid state nanocavity. However, we emphasize that such four-level systems could in principle also be implemented in solid state. As the system is driven by two classical fields, we monitor the second-order autocorrelation function of the field inside the cavity to study the strength of the photon blockade achievable in a cavity with mode volume comparable to cubed wavelength of the probe light.





  \begin{figure}[t]
   \begin{center}
   \begin{tabular}{c}
     \includegraphics[width=9cm]{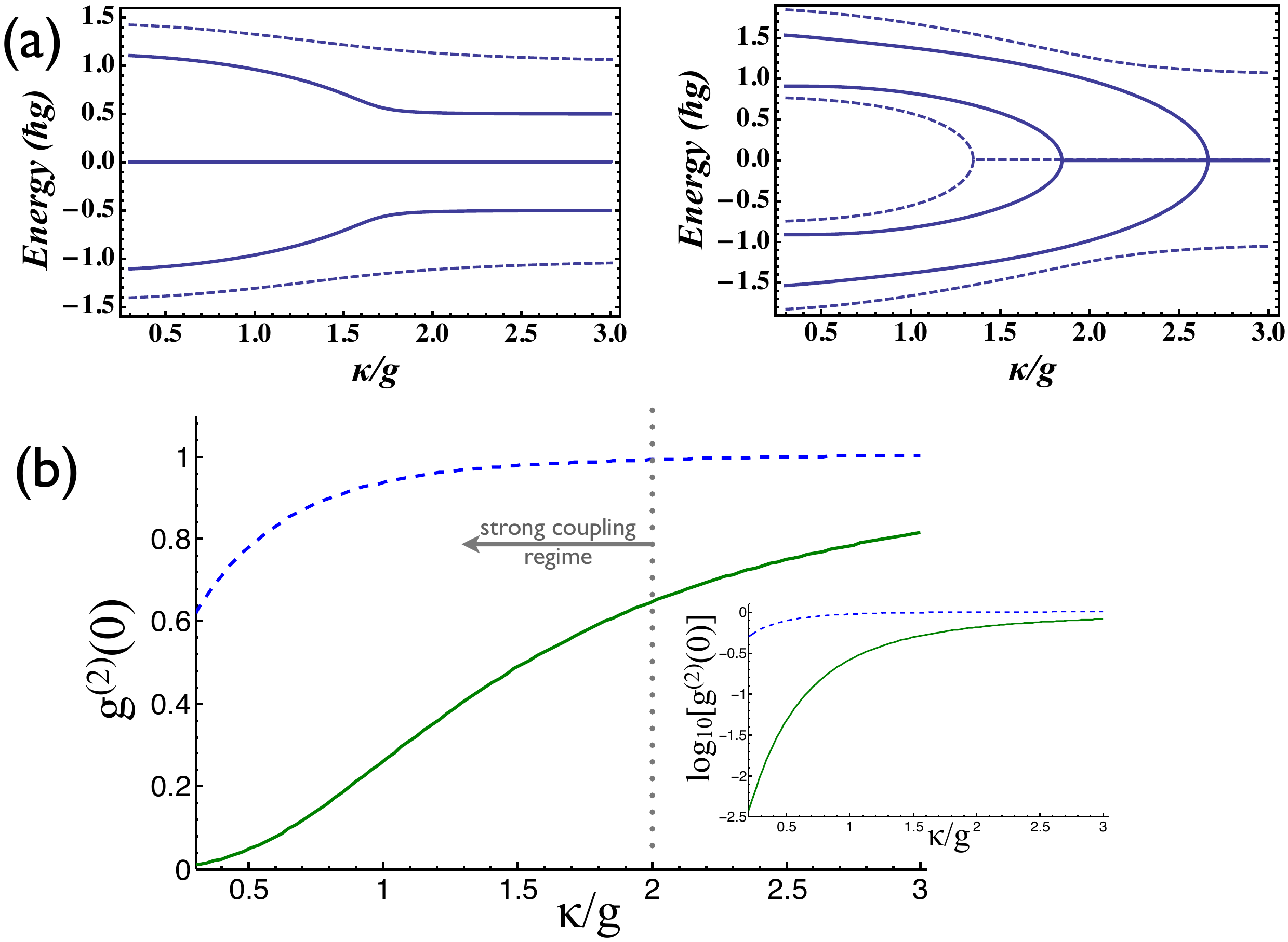}
   \end{tabular}
   \end{center}
   \caption{\label{fig2} (Color online) (a) Energy splitting of the levels in the first (left) and second (right) manifold of the dressed states for an idealized, i.e. resonant, system with $g_{eg}=g_{fs}=g$. Here, $\Omega=g_{eg}/2$ (solid lines) and $\Omega=g_{eg}$ (dashed lines). $\gamma_e/g=\gamma_f/g=3\times10^{-3}$, $\gamma_s/g= 3\times10^{-4}$. (b) Numerical simulation of the second order correlation function $g^{(2)}(0)$ of the light transmitted through such ideal system for $g_{eg}/2\pi=3$~GHz and $\Omega=g_{eg}$ (solid green curve) compared to that of a two-level system with the same parameters (dashed blue curve). The inset plots $g^{(2)}(0)$ on a log scale. For both simulations, $\mathcal{E}/2\pi=0.1$~GHz and the decay rates of the excited states are $2\pi \times 10$~MHz.}      
\end{figure} 

\section{Basic model}  
The Hamiltonian describing the coherent dynamics of a system depicted in Fig.\ref{fig1}(a) consisting of a four-level emitter and a cavity is given by \cite{Werner1999, Rebic1999}
\begin{equation}
H=H_o+H_{int}+H_{drive},
\end{equation}
where, assuming the rotating wave approximation, 

\begin{eqnarray}
H_o&=&\omega_c a^{\dag}a + \omega_{sg}\vert s \rangle \langle s \vert+ \omega_{eg}\vert e \rangle \langle e \vert +\omega_{fg}\vert f \rangle \langle f \vert \nonumber \\
H_{int}&=& g_{eg}(a^{\dag}\vert g\rangle \langle e\vert+a\vert e\rangle \langle g\vert)+g_{fs}(a^{\dag}\vert s\rangle \langle f\vert+a\vert f\rangle \langle s\vert) \nonumber \\
            &+&\Omega e^{-i\omega_d t}\vert e\rangle \langle s\vert +\Omega^* e^{i\omega_d t}\vert s\rangle \langle e\vert \nonumber \\
H_{drive}&=&\mathcal{E}e^{-i\omega_p t}a^{\dag}+\mathcal{E}^*e^{i\omega_p t}a \nonumber.
\end{eqnarray}
In this description, $a$ is the annihilation operator for the cavity mode, and the classical weak probe field couples to the cavity at a rate  $\mathcal{E}$. Another classical field with polarization orthogonal to the polarization of the cavity mode provides coupling between levels  $\vert s\rangle$ and $\vert e\rangle$ with Rabi frequency $\Omega$ without directly injecting photons into the cavity mode. The field of the cavity mode couples levels $\vert g\rangle$ and $\vert e\rangle$ and levels $\vert s\rangle$ and $\vert f\rangle$ with rates $g_{eg}$ and $g_{fs}$, respectively. In the above equations, $\omega_c$ is the frequency of the cavity mode and $\omega_{ij}$ are the frequencies of the atomic transition from level $\vert j\rangle$ to level $\vert i \rangle$  in Fig. \ref{fig1}(a) and $\hbar$ is set to 1. This analysis is relevant for any single-mode cavity, and does not involve any typical characteristics of a nanocavity.
By transforming into a rotating frame, the original Hamiltonian will change into 
\begin{equation}
\label{rotham}
\mathcal{H}=U^{\dag}HU + i{dU^{\dag} \over dt}U=\mathcal{H}_o+\mathcal{H}_{int}+\mathcal{H}_{drive}, 
\end{equation}
with
\begin{eqnarray}
U&=& e^{-i[\omega_p(a^{\dag}a+\vert e\rangle \langle e \vert)+(\omega_p-\omega_d)\vert s\rangle \langle s \vert+(2\omega_p-\omega_d)\vert f \rangle \langle f \vert]}\nonumber \\
\mathcal{H}_o&=&-\Delta_c a^{\dag}a -\Delta_{sg}\vert s \rangle \langle s \vert-\Delta_{eg}\vert e \rangle \langle e \vert -\Delta_{fg}\vert f \rangle \langle f \vert \nonumber\\ 
\mathcal{H}_{int}&=& g_{eg}(a^{\dag}\vert g\rangle \langle e\vert+a\vert e\rangle \langle g\vert)+g_{fs}(a^{\dag}\vert s\rangle \langle f \vert+a\vert f\rangle \langle s\vert) \nonumber \\
            &+&\Omega \vert e\rangle \langle s\vert +\Omega^* \vert s\rangle \langle e\vert \nonumber \\
\mathcal{H}_{drive}&=&\mathcal{E}a^{\dag}+\mathcal{E}^*a \nonumber,
\end{eqnarray}
where $\Delta_c=\omega_p-\omega_c$,  $\Delta_{sg}=\omega_p-\omega_{sg}-\omega_d$, $\Delta_{eg}=\omega_p-\omega_{eg}$, $\Delta_{fg}=2\omega_p-\omega_d-\omega_{fg}$.
In the limit of a very weak probe field ($\mathcal{E}\rightarrow 0$), the couplings between the bare states of the lowest manifolds of the overall system are depicted in Fig. \ref{fig1} (b), where each bare state is described by the state of the atom and the number of photons present in the cavity mode. The new eigenstates of the system are sketched in Fig. \ref{fig1} (c) for the resonant case ($\Delta_c=\Delta_{sg}=\Delta_{eg}=\Delta_{fg}=0$). The number of these dressed states in each manifold provides a basic explanation of the mechanism behind the photon blockade in this system. Specifically, if the probe is resonant with the bare cavity, the first photon will be resonant to the $\vert g, 0\rangle \rightarrow \vert D_1, 0\rangle$ transition, while the second photon will be out of resonance with any transition available out of state  $\vert D_1, 0\rangle$ to states $\vert D_2, j\rangle_{j\in \{\pm2,\pm1\}}$  of the second manifold \cite{Werner1999,Rebic1999}.      
If we take into account the dissipation of the cavity field to the environment with a decay rate $\kappa$ and the spontaneous decay rates between the atomic levels $\gamma_{sg}$, $\gamma_{eg}$, $\gamma_{es}$, $\gamma_{fg}$, and $\gamma_{fs}$, we can describe the dynamics of system with a master equation
\begin{equation}
\label{master}
{d\rho\over dt}=-{i\over \hbar}[\mathcal{H}, \rho]+2\kappa\mathcal{L}[a]+2\sum_{ij}\gamma_{ij}\mathcal{L}[\sigma_{ij}],
\end{equation}
where $\rho$ is the system's density matrix, $ij \in \{sg, eg, es, fg, fs\}$ and $\sigma_{ij}=\vert j \rangle \langle i\vert$. $\mathcal{L}[D]$ is the Lindblad superoperator on operator $D$ used to model the incoherent decays and is given by
\begin{equation}
\mathcal{L}[D]=D\rho D^{\dag} - {1\over 2}D^{\dag}D\rho-{1\over 2} \rho D^{\dag}D.
\end{equation}
For simplicity, we will ignore additional decay mechanisms and set $\gamma_s=\gamma_{sg}$, $\gamma_e=\gamma_{eg} + \gamma_{es}$, and $\gamma_f=\gamma_{fg}+\gamma_{fs}$ (Fig. \ref{fig1} (a)). 
We numerically simulate the dynamics of the system described by equation (\ref{master}) using the routines provided in the quantum optics toolbox \cite{qotoolbox} with up to six-photon Fock states.

\begin{figure}[t]
   \begin{center}
   \begin{tabular}{c}
     \includegraphics[width=6cm]{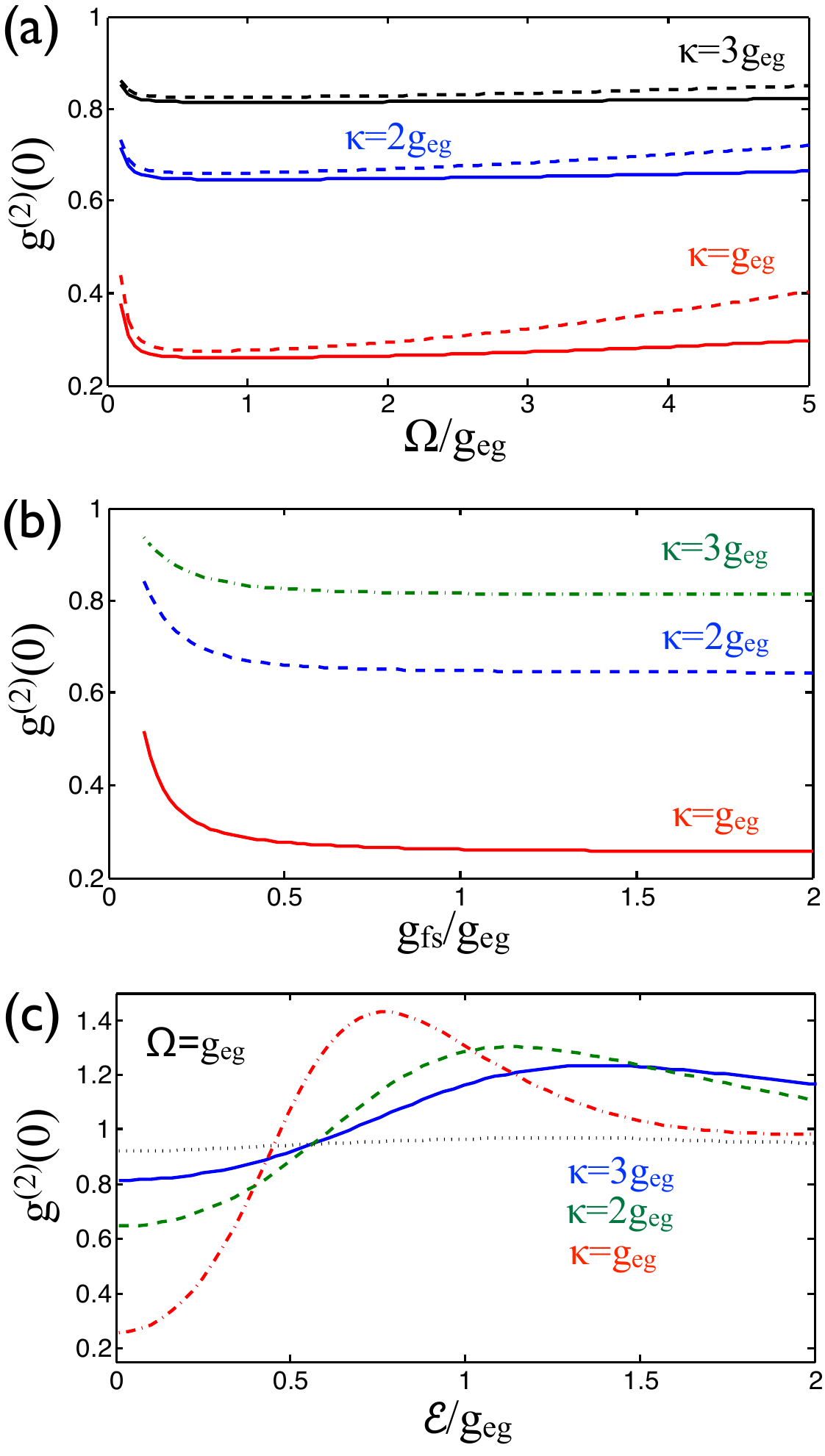}
   \end{tabular}
   \end{center}
   \caption{\label{fig3} (Color online) (a) A numerical simulation of the second order correlation function $g^{(2)}(0)$ of the light transmitted through the resonant system as a function of the amplitude of the classical field $\Omega$ with respect to $g_{eg}$ for different values of $\kappa$. Here, $g_{eg}/2\pi=3$~GHz,  $\mathcal{E}/2\pi=0.1$~GHz, $g_{fs}=g_{eg} $ (solid curves), $g_{fs}=g_{eg}/2 $ (dashed curves). (b) Numerically simulated $g^{(2)}(0)$ as a function of $g_{fs}/g_{eg}$ for $g_{eg}/2\pi=3$~GHz,  $\mathcal{E}/2\pi=0.1$~GHz, and $\Omega=g_{eg}/2$.  (c) Numerically simulated $g^{(2)}(0)$ as a function of the amplitude of the probe field $\mathcal{E}$ for $g_{eg}/2\pi=3$~GHz,  $\Omega=g_{eg}$. The value of $\kappa$ in this simulation is  $3g_{eg}$ (solid blue curve),  $2g_{eg}$ (dashed green curve), and $g_{eg}$ (dash-dot red curve). For comparison, the dotted black curve plots $g^{(2)}(0)$ as a function of $\mathcal{E}$ for photon blockade with a two level quantum emitter in a cavity with $\kappa=g_{eg}$. In all three parts of the figure, $\gamma_{eg}=\gamma_{es}=\gamma_{fg}=\gamma_{fs}=2\pi \times10$~MHz and $\gamma_{sg}=2\pi \times1$~MHz. }      
\end{figure} 

\section{Fully resonant case}
We start by looking at the idealized case of a fully resonant system ($\Delta_c=\Delta_{sg}=\Delta_{eg}=\Delta_{fg}=0$). This fully resonant case has been explored in detail in previous works \cite{Werner1999,Rebic1999, Greentree2000, Rebic2002a, Rebic2002b} with parameters corresponding to atomic cQED experiments in Fabry-Perot cavities ($g_{eg}/2\pi= 120$~MHz) \cite{Hood1998} in the regime of strong coupling ($g_{eg}>5\kappa$). Here, we revisit the resonant case but perform our numerical simulations with parameters that can be expected for a system based on a photonic crystal microcavity, in particular $g_{eg}/2\pi =3$~GHz. This is a fairly conservative value for atom-photon coupling achievable in a photonic-crystal nano-cavity, as values of $2\pi \times 17$~GHz have been predicted by Lev \textit{et al.} \cite{Lev2004} for closed transitions in cesium. We choose the lower value to budget for experimental imperfections, such as the possible use of cavity designs with larger mode volume, the use of weaker transitions to implement the four level emitter in a realistic atom, and the atom not being localized in the cavity field maximum, perhaps due to the atom's thermal motion or because the atom is trapped above the surface of the photonic crystal. For the cavity field decay, we focus on the interval of  $\kappa / 2\pi \approx 1 - 10$~GHz, corresponding to quality factors of $Q \sim 3\times 10^4$ to $3 \times 10^5$. For the chosen value of $g_{eg}$, this range includes the transition point $g_{eg}\approx{\kappa \over 2}$ between the weak and strong atom-field coupling in the cavity.  Values from this range have been observed in solid state cQED experiments in GaAs cavities \cite{Faraon2008a, Hennessy2007}, and values  of Q even higher have been predicted for air-mode cavities fabricated in other materials \cite{Quan2011}.  Particular materials of interest would include GaP \cite{Rivoire2008}, SiN, GaN, and SiO$_2$ \cite{Gong2010}, as these -- unlike GaAs -- remain transparent for light at wavelengths corresponding to commonly used optical transitions of alkali metals.  


The energy splittings of the dressed states in the first and second manifold are plotted as a function of the cavity field decay $\kappa$ in Fig. \ref{fig2} (a). These plots were obtained by numerically evaluating and then plotting the real part of the eigenvalues of the Hamiltonian $\mathcal{H}$ from equation (\ref{rotham}) with complex detunings to include the decay mechanisms \cite{Greentree2000}. For the chosen parameters, the first manifold is symmetrically split  into three levels, with one of the levels remaining at the energy of the original bare cavity. Qualitatively, the behavior of the first manifold is relatively insensitive to the value of $\kappa$ and $\Omega$. On the other hand, the second manifold splits into three and eventually into four distinct energy levels with a decreasing value of $\kappa$, while the value of $\kappa$ at which these splits occur can be affected by changing the ratio between $g_{eg}$ and $\Omega$. A detailed calculation of the exact energies of the dressed states can be found in reference \cite{Rebic2002a}.

The effects of this level structure on the photon statistics inside the cavity can be seen in Fig. \ref{fig2} (b), where we plot the second order correlation function $g^{(2)}(0)$ of the field inside the cavity as a function of the $g/ \kappa$ ratio with the probe field resonant with the bare cavity ($\Delta_c =0$, green curve).  We see that,  in a fashion similar to schemes based on a two-level emitter coupled to a bimodal cavity \cite{Kimble2008, Majumdar2012}, 
the photons are significantly anti-bunched ($g^{(2)}(0) \approx 0.6$) already for $g_{eg} =\kappa/2$, and the value of $g^{(2)}(0)$ for this system remains appreciably below what can be achieved with a two-level emitter coupled to a single mode cavity (blue dashed curve) even as $\kappa$ increases.

This is particularly interesting when compared to the energy splittings of the dressed states plotted in Fig. \ref{fig2} (a), since one would not expect a blockade until the energy  degeneracy of the bare states is lifted by strong enough emitter-field coupling.
For comparison, we also plot the $g^{(2)}(0)$ expected for a two-level atom coupled to a cavity with identical parameters and probe detuning optimized for this system (blue curve, $\Delta_c \approx 1.5g$). The anti-bunching resulting from a four-level system is much stronger than that achievable in a two-level system, with the difference exceeding two orders of magnitude for  $g/ \kappa>5$ (Fig. \ref{fig2}(b), inset). This is similar to predictions made based on parameters achievable in a Fabry-Perrot cavity \cite{Werner1999}.      
 \begin{figure}[b]
   \begin{center}
   \begin{tabular}{c}
     \includegraphics[width=8.5cm]{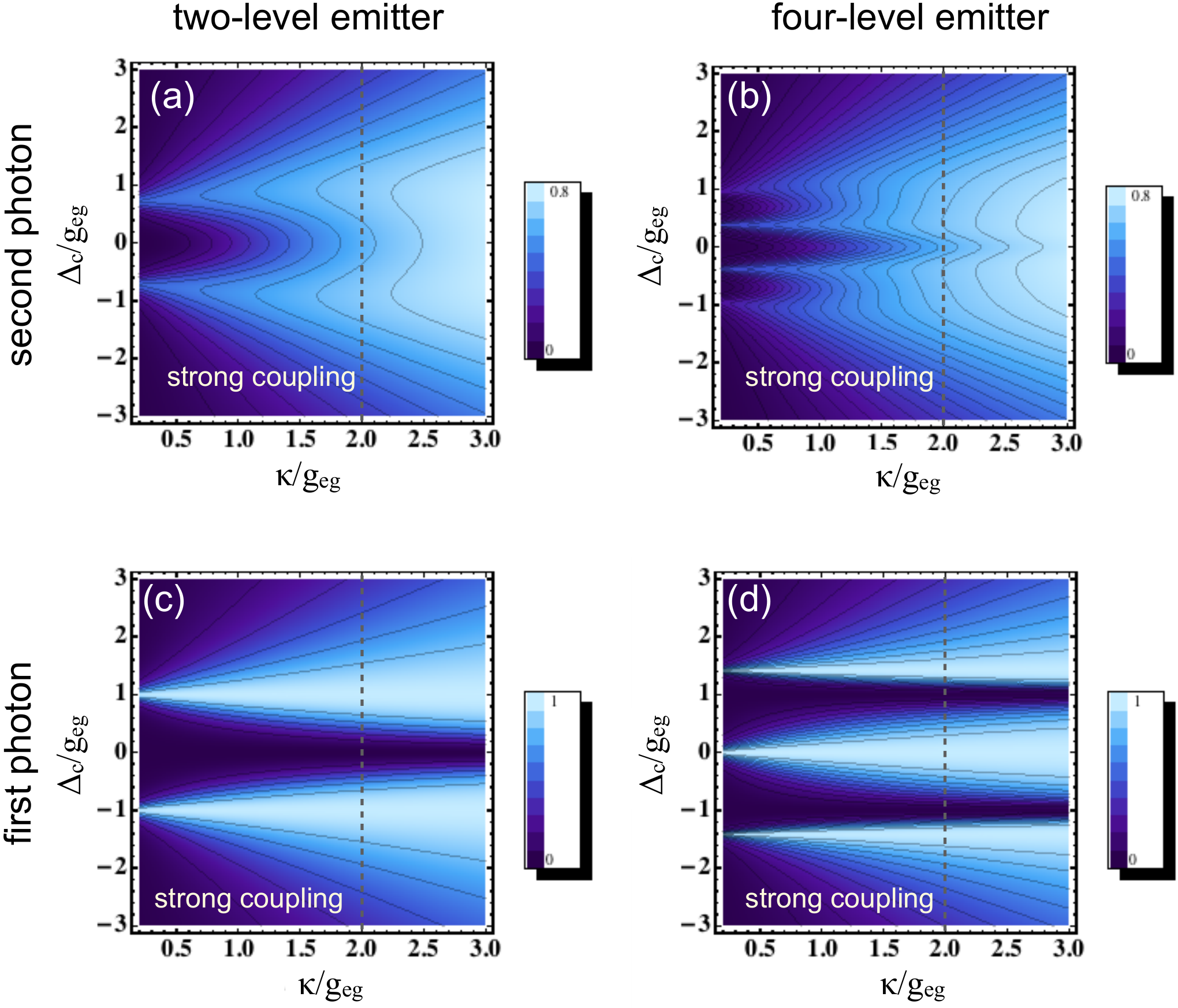}
   \end{tabular}
   \end{center}
   \caption{\label{fig-int1} Estimated transmission of the first two photons through a cavity with a resonant quantum emitter as a function of probe detuning: (a) and (c) for the case of a two-level emitter ($\Delta_{eg}= \Delta_c$); (b) and (d) for an ideal four-level emitter ($\Delta_{eg} =\Delta_{fs}= \Delta_c$) with parameters from Fig \ref{fig2} (b). The border of the strong coupling regime for the $\vert g\rangle \rightarrow \vert e \rangle$ transition is marked by the vertical dashed line in each plot.}
\end{figure} 
In Fig. \ref{fig3}, we numerically simulate the effects of the system's key parameters on the photon statistics at the onset ($\kappa=2g_{eg}$), within ($\kappa=g_{eg}$), and outside of ($\kappa=3g_{eg}$) the strongly coupled regime. Note that we take $g_{eg}\approx \kappa/2$ as the onset of strongly coupled regime, as this is when the energies of the eigenstates of the first rung of Jaynes-Cummings ladder for a system consisting of a two-level emitter coupled to a cavity become non-degenerate \cite{Englund2007}. For a given value of $\kappa$, the resulting $g^{(2)}(0)$ is fairly robust against variations in the magnitude of the coherent field $\Omega$  (Fig. \ref{fig3} (a)). Similarly, $g^{(2)}(0)$ remains nearly constant for $g_{fs}>g_{eg}/2$ (Fig. \ref{fig3} (b)). Lastly, the light inside the cavity remains anti-bunched for a broad range of values of the probe field amplitude (Fig. \ref{fig3} (c)), before changing into the bunching regime predicted in \cite{Rebic2002b}.   
 
 \section{The intuitive picture}
 An intuitive picture of why a significantly better photon blockade is expected with a four-level emitter and why this blockade also happens outside of the strong coupling regime -- i.e. before the energy degeneracy of the dressed states is lifted as shown in Fig.  \ref{fig2} (a ) --  can be provided by comparing the transmission of the system for the first photon and for the second photon coupled into the cavity. An estimate of these transmissions can be obtained by taking the non-Hermitian Hamiltonians characterizing the system's transition between its ground and first, as well as first and second manifolds, applying these Hamiltonians to wave-functions describing the system state in these manifolds, and then finding a steady-state solution under the condition of a weak probe field.

The non-Hermitian Hamiltonian describing the system's transition between the ground state and the first manifold can be written as   
\begin{eqnarray}
\label{int1}   
  \tilde{\mathcal{H}}^{(1)}=&-&(i \kappa+\Delta_c) a^{\dag}a -(i \gamma_{sg}+\Delta_{sg})\vert s \rangle \langle s \vert \nonumber\\
  &-&(i \gamma_{eg}+\Delta_{eg})\vert e \rangle \langle e \vert  + g_{eg}(a^{\dag}\vert g\rangle \langle e\vert+a\vert e\rangle \langle g\vert) \nonumber \\
  &+&\Omega \vert e\rangle \langle s\vert +\Omega^* \vert s\rangle \langle e\vert +\mathcal{E}a^{\dag}+\mathcal{E}^*a.
  \end{eqnarray} 
  If we assume the system is in a state described by
  \begin{equation}
  \label{int1-wf}
  \vert \psi^{(1)}\rangle=c^{(1)}_o \vert g, 0\rangle +c^{(1)}_1 \vert g, 1\rangle+c^{(1)}_2\vert e, 0\rangle+c^{(1)}_3 \vert s, 0\rangle,
  \end{equation}
then the equation we are trying to solve is 
\begin{equation}
\label{int1-se} 
i {d\over dt} \vert \psi^{(1)}\rangle= \tilde{\mathcal{H}}^{(1)}\vert \psi^{(1)}\rangle.
\end{equation}
 In the limit of $\mathcal{E}\rightarrow0$, we set $c^{(1)}_o\approx1$ and we neglect the manifold-connecting effects of the probe, such as the $\vert g, 1\rangle \rightarrow \vert g, 2\rangle$,  $\vert e, 0\rangle \rightarrow \vert e, 1\rangle$, $\vert s, 0\rangle \rightarrow \vert s, 1\rangle$ coupling. We define the transmission of the first photon though the system as 
 \begin{equation} 
 T^{(1)}={\langle a^{\dag}a \rangle \over{\langle a^{\dag}a \rangle^{(1)}_o}}\approx{\vert c^{(1)}_1\vert^2\over \vert \mathcal{E}\vert^2}\kappa^2,
 \end{equation}
 with  $\langle a^{\dag}a \rangle^{(1)}_o$ being the transmission of the first photon though an empty cavity on resonance, and solve equation (\ref{int1-se}) in steady state (${d\over dt}c^{(1)}_j=0$) to find the value of  $c^{(1)}_1$.
 
 In a similar fashion,  the non-Hermitian Hamiltonian describing the system's transition between the first and second manifold can be written as   
\begin{equation}
\label{int2}   
  \tilde{\mathcal{H}}^{(2)}=  \tilde{\mathcal{H}}^{(1)} -(i \gamma_{fs}+\Delta_{fs})\vert f \rangle \langle f \vert
    + g_{fs}(a^{\dag}\vert s\rangle \langle f\vert+a\vert f\rangle \langle s\vert).
  \end{equation} 
We assume the system is in a state described by
  \begin{equation}
  \label{int2-wf}
  \vert \psi^{(2)}\rangle=c^{(2)}_o \vert g, 1\rangle +c^{(2)}_1 \vert g, 2\rangle+c^{(2)}_2\vert e, 1\rangle+c^{(2)}_3 \vert s, 1\rangle+c^{(2)}_4 \vert f, 0\rangle
  \end{equation}
  and we are trying to extract the information about the transmission encountered by the second photon (assuming the first photon has already coupled into the system) by solving the equation 
  \begin{equation}
\label{int2-se} 
i {d\over dt} \vert \psi^{(2)}\rangle= \tilde{\mathcal{H}}^{(2)}\vert \psi^{(2)}\rangle.
\end{equation}
Again, in the limit of $\mathcal{E}\rightarrow0$, we set $c^{(2)}_o\approx1$, neglect the manifold-connecting effects of the probe, define the transmission of the second photon as 
\begin{equation} 
 T^{(2)}={\langle a^{\dag}a \rangle \over{\langle a^{\dag}a \rangle^{(2)}_o}}\approx2{\vert c^{(2)}_1\vert^2\over \vert \mathcal{E}\vert^2}\kappa^2,
 \end{equation}
 with  $\langle a^{\dag}a \rangle^{(2)}_o$ being the transmission of the first photon though an empty cavity on resonance, and solve equation (\ref{int2-se}) in steady state to find the value of  $c^{(2)}_1$. Note that we only included state $\vert g, 1\rangle$ from the first manifold in Eq. (\ref{int2-wf}) as the other states will not have a leading-order effect on the transmission of the second photon. 
  
 An analogous approach can be used to estimate the transmission of the first and second photon through a cavity with a two-level emitter. The result of that calculation is plotted in Fig. \ref{fig-int1} (a) and (c) for comparison with the four-level emitter case. We see that deep in the strongly coupled regime ($\kappa = g_{eg}/3$ in Fig. \ref{int1} (c)) the best transmission of the first photon is achieved for $\vert {\Delta_{c}\over g_{eg}}\vert=1$, while the transmission of the second photon at this detuning is significantly suppressed (Fig. \ref{fig-int1} (a)). Unfortunately, with increasing cavity field decay $\kappa$, the transmission of the second photon increases for $\vert \Delta_{c}\vert =g_{eg}$. One can remain in the photon blockade regime by adjusting the frequency of the probe -- effectively trading off the transmission of the first photon for reduced transmission of the second photon -- until the transmission peaks of the second photon become too broad and adjusting $\Delta_c$ will not have a noteworthy effect on the statistics of the transmitted light anymore.

 
Returning to the ideal four-level emitter, we see that when the probe is resonant with the bare cavity ($\Delta_c=0$, Fig. \ref{fig-int1} (d)) the transmission of the first photon is maximized, while at the same time there is a minimum in the transmission of the second photon (Fig. \ref{fig-int1} (b)). This convenient frequency alignment of the maximum transmission for the first photon with a local minimum in the transmission of the second photon is present even outside the strongly coupled regime and while the energies of the dressed states (Fig. \ref{fig2}(a)) remain degenerate. This minimum can be thought of as a generalized case of dipole induced transparency (DIT) \cite{Waks2006} and a similar minimum in the transmission of the second photon at $\Delta_c=0$ can be observed in the case of a two-level emitter (Fig. \ref{fig-int1} (a)). In this case though, this minimum coincides with a minimum in the transmission of the first photon (Fig. \ref{fig-int1} (c)) and is therefore not useful for implementing photon blockade. 

To summarize, the optimal frequency alignment of the maximum transmission for the first photon with a local minimum in the transmission of the second photon provides an intuitive explanation for the superior photon blockade performance of the four-level emitter.  At the same time, for both the two-level and the four-level emitter case, the blockade performance deteriorates with increasing $\kappa$ mainly because of broadening of the line-widths of the dressed states, which makes it easier for the second photon to off-resonantly couple into the system.    
  
\section{Off-resonant system}
\begin{figure}[t]
   \begin{center}
   \begin{tabular}{c}
     \includegraphics[width=8.5cm]{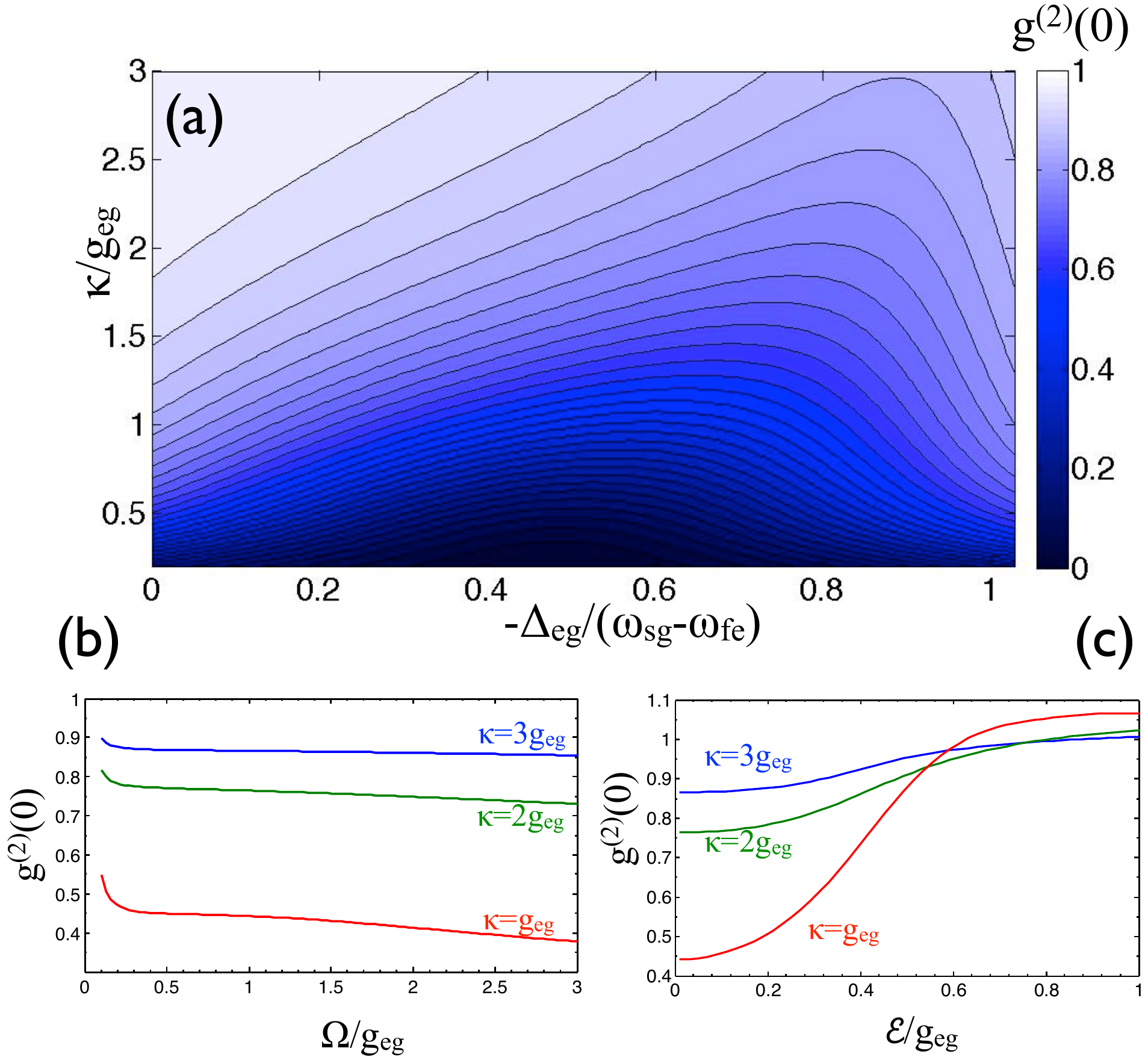}
   \end{tabular}
   \end{center}
   \caption{\label{fig4} (Color online) (a) Numerical simulation of $g^{(2)}(0)$ as a function of cavity decay rate $\kappa$ and probe detuning $\Delta_{eg}$ from the $\vert g\rangle \rightarrow \vert e \rangle$ transition in a four-level system based on $^{133}$Cs ($\omega_{sg}-\omega_{fe}=2\pi \times 8.941$~GHz). Here, the probe is resonant with the cavity ($\Delta_c=0$) and we keep $\Delta_{sg}=0$ by setting $\omega_d=\omega_p-\omega_{sg}$. $g_{eg}/2\pi=3$~GHz, $g_{fs}=1.5g_{eg}$, $\Omega=g_{eg}$. (b) $g^{(2)}(0)$ as a function of the amplitude of the classical field $\Omega$ for different values of $\kappa$. (c) Effects of the probe field amplitude $\mathcal{E}$ on $g^{(2)}(0)$. In both (b) and (c) the value of $\Delta_{eg}$ is chosen to minimize $g^{(2)}(0)$. In all plots, $\gamma_{eg}=\gamma_{es}=\gamma_{fg}=\gamma_{fs}=2\pi \times10$~MHz and $\gamma_{sg}=2\pi \times1$~MHz.}      
\end{figure} 
 
 \begin{figure}[b]
   \begin{center}
   \begin{tabular}{c}
     \includegraphics[width=8.5cm]{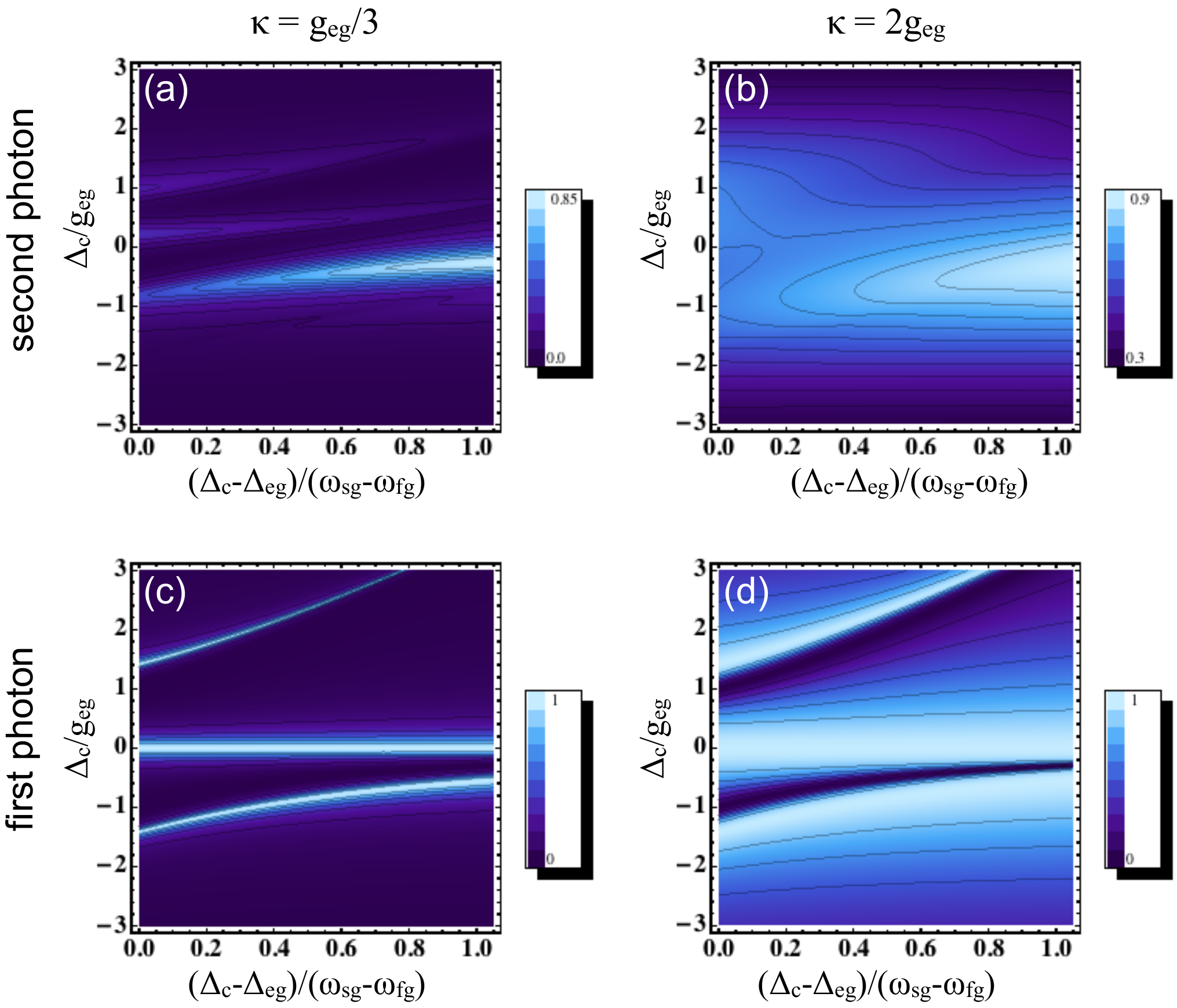}
   \end{tabular}
   \end{center}
   \caption{\label{fig-int2} Estimated transmission of the first two photons through the system numerically simulated in Fig. \ref{fig4} as a function of detuning $\Delta_c$ of the probe from the cavity: (a) and (c) for the case of strong coupling regime between the emitter and the cavity ($\kappa=g_{eg}/3$); (b) and (d) for the system being on the verge of strong coupling regime ($\kappa=2g_{eg}$).}      
\end{figure} 
While the fully-resonant system is interesting as a model for understanding the mechanisms behind the photon blockade and for estimating the limits of its behavior, a practical implementation of this scheme will most likely have to be done with a four-level quantum emitter in which the frequencies of the $\vert g\rangle \rightarrow \vert e \rangle$ and $\vert s\rangle \rightarrow \vert f \rangle$ transitions differ significantly. For example, in the alkali atoms used commonly for cQED experiments, such as rubidium or cesium, the frequency difference between these two transitions is about two orders of magnitude larger than the atom-field coupling rate $g$ achieved in current experimental atomic cQED systems \cite{Kimble2005, Kimble2008, Koch2011}, which report $g/2\pi\sim10 - 100$~MHz. The only exceptions are disk cavities reported by Barclay \textit{et al.}  \cite{Barclay2006}, who predict $g/2\pi\approx 0.9$~GHz for atoms displaced $100$~nm from the cavity surface. However, coupling of atoms to these cavities has not yet been demonstrated to the best of our knowledge.

In this section, we therefore base the four-level quantum emitter on the $^{133}$Cs atom, in which $\omega_{sg}-\omega_{fe}\approx2\pi \times 8.941$~GHz, to study the photon blockade under conditions that are closer to a realistic system. Of course this representation of the cesium atom, while maintaining the frequency scales, is still highly simplified. To study this system in detail one needs to implement a full simulation including the effects of all participating levels of the cesium atom in a fashion similar to the work presented by Birnbaum \textit{et al.} \cite{Birnbaum2006}.

Following the approach suggested by Greentree \textit{et al.} \cite{Greentree2000}, we keep the probe resonant with the cavity, adjust $\omega_d$ to stay on resonance with the two-photon transition  $\vert g\rangle \rightarrow \vert s \rangle$ ($\omega_d=\omega_p-\omega_{sg}$), and vary the probe detuning $\Delta_{eg}$ from the $\vert g\rangle \rightarrow \vert e \rangle$ transition. In this approach, the probe can couple well into the cavity, while adjusting $\Delta_{eg}$ optimizes the strength of the interaction of the cavity photons with the two cavity-coupled transitions $\vert g\rangle \rightarrow \vert e \rangle$ and $\vert s\rangle \rightarrow \vert f \rangle$.
Figure \ref{fig4}(a) then plots $g^{(2)}(0)$ as a function of cavity decay rate $\kappa$ and probe detuning $\Delta_{eg}$ from the $\vert g\rangle \rightarrow \vert e \rangle$ transition for a probe resonant with the cavity ($\Delta_c=0$). We see that for each value of $\kappa$ within the range under consideration, there is a value of $\Delta_{eg}$ optimizing the anti-bunching of the light inside the cavity (i.e., $g^{(2)}(0)$ is minimized). Similarly, as in the case of the ideal atom, changing the value of $\Omega$ affects blockade only slightly (Fig. \ref {fig4}(b)) and the light inside the cavity remains anti-bunched for a relatively wide range of intensities of the probe light (Fig. \ref {fig4}(c)). Note that both of these plots display the minimum value of $g^{(2)}(0)$ for a given $\Omega$ or $\mathcal{E}$ that can be achieved by adjusting $\Delta_{eg}$ while keeping the probe resonant with the cavity.

The effects of adjusting $\Delta_{eg}$ on the photon blockade in an off-resonant system can also be observed in the intuitive picture described in the previous section.  Figure \ref{fig-int2} plots the estimated transmission for the first and second photon through the system numerically simulated in Fig. \ref{fig4} (a) for strong coupling regime ($\kappa=g_{eg}/3$) and outside of strong coupling regime ($\kappa=2g_{eg}$). In both cases we see that the transmission of the first photon remains highest at $\Delta_c=0$ as the detuning of the probe $\Delta_{eg}$ from the $\vert g\rangle \rightarrow \vert e \rangle$ transition is changed (Fig. \ref{fig-int2} (c), (d)). On the other hand, there is a minimum with respect to $\Delta_{eg}$ in the transmission of the second photon when $\Delta_c=0$. This minimum is particularly pronounced in the strong coupling regime (Fig. \ref{fig-int2} (a)) and remains present, although in a washed-out form, even as the cavity decay is increased to $\kappa=2g_{eg}$ (Fig. \ref{fig-int2} (b)). However, in terms of the value of $\Delta_{eg}$ at which the minimum happens, the agreement with the numerical simulation in Fig. \ref{fig4} is only qualitative, most likely due to the manifold-connecting effects of the probe field $\mathcal{E}$ that are neglected in the intuitive picture. 

\begin{figure}
   \begin{center}
   \begin{tabular}{c}
     \includegraphics[width=8.5cm]{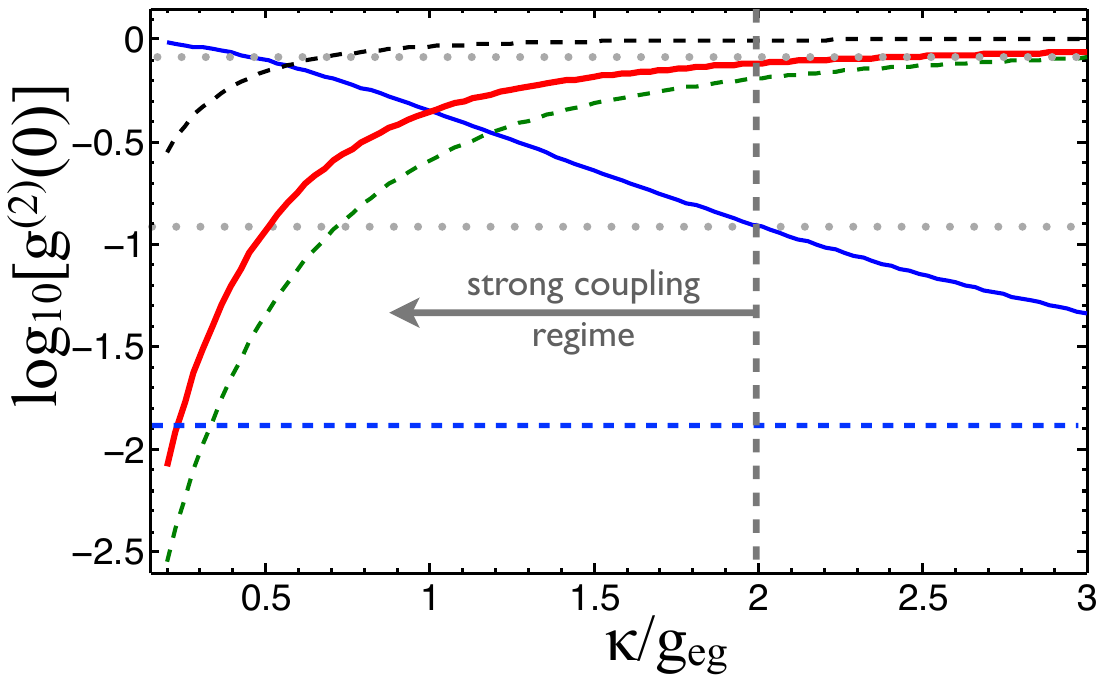}
   \end{tabular}
   \end{center}
   \caption{\label{fig5} (Color online) Comparison of $g^{(2)}(0)$ achievable with a four-level off-resonant emitter based on cesium (solid red curve) with a resonant four-level emitter from Fig. \ref{fig2}(c) (dashed green), with a comparable two-level emitter (dashed black curve), and with a two-level atom coupled to a bimodal cavity (solid blue curve and dashed blue horizontal line). The two dotted horizontal grey lines correspond to blockade with a four-level off-resonant emitter based on cesium and parameters demonstrated in current cQED experiments: $g_{eg}/2\pi =120$~MHz, $\kappa/2\pi=40$~MHz and $4$~MHz for the upper and lower grey line, respectively. The decay rates of the atomic excited states are equal for all plots.}      
\end{figure} 

The photon blockade achievable with the cesium-based four-level atom is put into perspective in Fig. \ref{fig5}. The solid red curve plots the minimum value of $g^{(2)}(0)$ with respect to $\Delta_{eg}$ from Fig. \ref{fig4}(a) as a function of $\kappa$. We see that the blockade is slightly worse than that achievable with the ideal resonant system (green dashed curve), but it still vastly outperforms the blockade achievable in an identical cavity with a comparable two level atom (black dashed curve). Note that we do not observe the photon bunching effects discussed in \cite{Chang2007, Rice1988}, as we believe they are not relevant in our parameter regime.
 For comparison, the two grey dotted lines mark the expected blockade with a four-level cesium-based atom and cavities with mode volume resulting in $g_{eg}/2\pi \approx 100$~MHz, which roughly represent  state of the art  cavities used in atomic cQED experiments, such as  \cite{Hood1998, Aoki2009, Alton2010}.  The upper line corresponds to $\kappa/2\pi=40$~MHz  achieved by Hood \textit{et al.} \cite{Hood1998}, while the lower line represents $\kappa/2\pi=4$~MHz, which is about five-times smaller than the field decay rate reported by Alton \textit{et al.} \cite{Alton2010}.    
Finally, the solid blue line represents the $g^{(2)}(0)$ expected with a two-level emitter coupled to two orthogonally polarized modes of a degenerate bimodal nanocavity as proposed by Majumdar \textit{et al.} \cite{Majumdar2012}, with the blue dashed line marking the lower limit of $g^{(2)}(0)$ in such system. We see that this proposal provides photon-blockade performance comparable to the four-level atom scheme analyzed here, but its implementation with atoms might be difficult due to the selection rules of atomic transitions. 
     
\section{Outlook}
In conclusion, our results show that, under conditions achievable with currently available photonic crystal nanocavities, the originally proposed photon blockade \cite{Imamoglu1997} should be experimentally observable with realistic four-level atoms and even without achieving the strong coupling regime between the atom and the cavity field. The potential for scalability of this scheme when implemented with cold atoms and photonic crystal nanocavities, as well as the robustness to variations in experimental parameters seen in the presented simulations, make the ``original" photon blockade a great candidate for experimental realization of networks of coupled nonlinear cavities in which the interactions between photons can be engineered \cite{Greentree2006}.
Additionally, the results presented here might be relevant for exploration of photon blockade with four-level solid state emitters coupled to nanocavities, such as nitrogen-vacancy centers in diamond \cite{Faraon2011a, Babinec2011, Englund2010} and charged quantum dots in a strong magnetic field \cite{Press2008}.

\section{Acknowledgements}
This work was supported by DARPA, grant number N66001-12-1-4011, and by the Air Force Office of Scientic Research, MURI Center for 
Multi-functional light-matter interfaces based on atoms and solids. A. R. was also supported by a Stanford Graduate Fellowship.

\bibliography{4-lev-bib}

\begin{thebibliography}{52}%
\makeatletter
\providecommand \@ifxundefined [1]{%
 \@ifx{#1\undefined}
}%
\providecommand \@ifnum [1]{%
 \ifnum #1\expandafter \@firstoftwo
 \else \expandafter \@secondoftwo
 \fi
}%
\providecommand \@ifx [1]{%
 \ifx #1\expandafter \@firstoftwo
 \else \expandafter \@secondoftwo
 \fi
}%
\providecommand \natexlab [1]{#1}%
\providecommand \enquote  [1]{``#1''}%
\providecommand \bibnamefont  [1]{#1}%
\providecommand \bibfnamefont [1]{#1}%
\providecommand \citenamefont [1]{#1}%
\providecommand \href@noop [0]{\@secondoftwo}%
\providecommand \href [0]{\begingroup \@sanitize@url \@href}%
\providecommand \@href[1]{\@@startlink{#1}\@@href}%
\providecommand \@@href[1]{\endgroup#1\@@endlink}%
\providecommand \@sanitize@url [0]{\catcode `\\12\catcode `\$12\catcode
  `\&12\catcode `\#12\catcode `\^12\catcode `\_12\catcode `\%12\relax}%
\providecommand \@@startlink[1]{}%
\providecommand \@@endlink[0]{}%
\providecommand \url  [0]{\begingroup\@sanitize@url \@url }%
\providecommand \@url [1]{\endgroup\@href {#1}{\urlprefix }}%
\providecommand \urlprefix  [0]{URL }%
\providecommand \Eprint [0]{\href }%
\@ifxundefined \urlstyle {%
  \providecommand \doi  [0]{\begingroup \@sanitize@url \@doi}%
  \providecommand \@doi [1]{\endgroup \@@startlink {\doibase
  #1}doi:\discretionary {}{}{}#1\@@endlink }%
}{%
  \providecommand \doi  [0]{doi:\discretionary{}{}{}\begingroup
  \urlstyle{rm}\Url }%
}%
\providecommand \doibase [0]{http://dx.doi.org/}%
\providecommand \Doi [0]{\begingroup \@sanitize@url \@Doi }%
\providecommand \@Doi  [1]{\endgroup\@@startlink{\doibase#1}\@@Doi}%
\providecommand \@@Doi [1]{#1\@@endlink}%
\providecommand \selectlanguage [0]{\@gobble}%
\providecommand \bibinfo  [0]{\@secondoftwo}%
\providecommand \bibfield  [0]{\@secondoftwo}%
\providecommand \translation [1]{[#1]}%
\providecommand \BibitemOpen [0]{}%
\providecommand \bibitemStop [0]{}%
\providecommand \bibitemNoStop [0]{.\EOS\space}%
\providecommand \EOS [0]{\spacefactor3000\relax}%
\providecommand \BibitemShut  [1]{\csname bibitem#1\endcsname}%
\bibitem [{\citenamefont {Imamo\u{g}lu}\ \emph {et~al.}(1997)\citenamefont
  {Imamo\u{g}lu}, \citenamefont {Schmidt}, \citenamefont {Woods},\ and\
  \citenamefont {Deutsch}}]{Imamoglu1997}%
  \BibitemOpen
  \bibfield  {author} {\bibinfo {author} {\bibfnamefont {A.}~\bibnamefont
  {Imamo\u{g}lu}}, \bibinfo {author} {\bibfnamefont {H.}~\bibnamefont
  {Schmidt}}, \bibinfo {author} {\bibfnamefont {G.}~\bibnamefont {Woods}}, \
  and\ \bibinfo {author} {\bibfnamefont {M.}~\bibnamefont {Deutsch}},\
  }\href@noop {} {\bibfield  {journal} {\bibinfo  {journal} {Physical Review
  Letters},\ }\textbf {\bibinfo {volume} {79}},\ \bibinfo {pages} {1467}
  (\bibinfo {year} {1997})}\BibitemShut {NoStop}%
\bibitem [{\citenamefont {Fulton}\ and\ \citenamefont
  {Dolan}(1987)}]{Fulton1987}%
  \BibitemOpen
  \bibfield  {author} {\bibinfo {author} {\bibfnamefont {T.~A.}\ \bibnamefont
  {Fulton}}\ and\ \bibinfo {author} {\bibfnamefont {G.~J.}\ \bibnamefont
  {Dolan}},\ }\href@noop {} {\bibfield  {journal} {\bibinfo  {journal} {Phys.
  Rev. Lett.},\ }\textbf {\bibinfo {volume} {59}},\ \bibinfo {pages} {109}
  (\bibinfo {year} {1987})}\BibitemShut {NoStop}%
\bibitem [{\citenamefont {Greentree}\ \emph {et~al.}(2006)\citenamefont
  {Greentree}, \citenamefont {Tahan}, \citenamefont {Cole},\ and\ \citenamefont
  {Hollenberg}}]{Greentree2006}%
  \BibitemOpen
  \bibfield  {author} {\bibinfo {author} {\bibfnamefont {A.~D.}\ \bibnamefont
  {Greentree}}, \bibinfo {author} {\bibfnamefont {C.}~\bibnamefont {Tahan}},
  \bibinfo {author} {\bibfnamefont {J.~H.}\ \bibnamefont {Cole}}, \ and\
  \bibinfo {author} {\bibfnamefont {L.~C.~L.}\ \bibnamefont {Hollenberg}},\
  }\href@noop {} {\bibfield  {journal} {\bibinfo  {journal} {Nature Phys.},\
  }\textbf {\bibinfo {volume} {2}},\ \bibinfo {pages} {856} (\bibinfo {year}
  {2006})}\BibitemShut {NoStop}%
\bibitem [{\citenamefont {Hartmann}\ \emph {et~al.}(2006)\citenamefont
  {Hartmann}, \citenamefont {Brand$\mathrm{\tilde{a}}$o},\ and\ \citenamefont
  {Plenio}}]{Hartmann2006}%
  \BibitemOpen
  \bibfield  {author} {\bibinfo {author} {\bibfnamefont {M.~J.}\ \bibnamefont
  {Hartmann}}, \bibinfo {author} {\bibfnamefont {F.~G.~S.}\ \bibnamefont
  {Brand$\mathrm{\tilde{a}}$o}}, \ and\ \bibinfo {author} {\bibfnamefont
  {M.~B.}\ \bibnamefont {Plenio}},\ }\href@noop {} {\bibfield  {journal}
  {\bibinfo  {journal} {Nature Phys.},\ }\textbf {\bibinfo {volume} {2}},\
  \bibinfo {pages} {849} (\bibinfo {year} {2006})}\BibitemShut {NoStop}%
\bibitem [{\citenamefont {Carusotto}\ and\ \citenamefont
  {Ciuti}(2012)}]{Carusotto2012}%
  \BibitemOpen
  \bibfield  {author} {\bibinfo {author} {\bibfnamefont {I.}~\bibnamefont
  {Carusotto}}\ and\ \bibinfo {author} {\bibfnamefont {C.}~\bibnamefont
  {Ciuti}},\ }\href@noop {} {\bibfield  {journal} {\bibinfo  {journal}
  {arXiv},\ \bibinfo {pages} {1205.6500}} (\bibinfo {year} {2012})}\BibitemShut
  {NoStop}%
\bibitem [{\citenamefont {Faraon}\ \emph {et~al.}(2008)\citenamefont {Faraon},
  \citenamefont {Fushman}, \citenamefont {Englund}, \citenamefont {Stoltz},
  \citenamefont {Petroff},\ and\ \citenamefont
  {Vu\v{c}kovi\'{c}}}]{Faraon2008a}%
  \BibitemOpen
  \bibfield  {author} {\bibinfo {author} {\bibfnamefont {A.}~\bibnamefont
  {Faraon}}, \bibinfo {author} {\bibfnamefont {I.}~\bibnamefont {Fushman}},
  \bibinfo {author} {\bibfnamefont {D.}~\bibnamefont {Englund}}, \bibinfo
  {author} {\bibfnamefont {N.}~\bibnamefont {Stoltz}}, \bibinfo {author}
  {\bibfnamefont {P.}~\bibnamefont {Petroff}}, \ and\ \bibinfo {author}
  {\bibfnamefont {J.}~\bibnamefont {Vu\v{c}kovi\'{c}}},\ }\href@noop {}
  {\bibfield  {journal} {\bibinfo  {journal} {Nat. Phys.},\ }\textbf {\bibinfo
  {volume} {4}},\ \bibinfo {pages} {859} (\bibinfo {year} {2008})}\BibitemShut
  {NoStop}%
\bibitem [{\citenamefont {Majumdar}\ \emph
  {et~al.}(2012){\natexlab{a}}\citenamefont {Majumdar}, \citenamefont
  {Bajcsy},\ and\ \citenamefont {Vu\v{c}kovi\'{c}}}]{Majumdar2012a}%
  \BibitemOpen
  \bibfield  {author} {\bibinfo {author} {\bibfnamefont {A.}~\bibnamefont
  {Majumdar}}, \bibinfo {author} {\bibfnamefont {M.}~\bibnamefont {Bajcsy}}, \
  and\ \bibinfo {author} {\bibfnamefont {J.}~\bibnamefont {Vu\v{c}kovi\'{c}}},\
  }\href@noop {} {\bibfield  {journal} {\bibinfo  {journal} {Phys. Rev. A},\
  }\textbf {\bibinfo {volume} {85}} (\bibinfo {year}
  {2012}{\natexlab{a}})}\BibitemShut {NoStop}%
\bibitem [{\citenamefont {Rosenblum}\ \emph {et~al.}(2011)\citenamefont
  {Rosenblum}, \citenamefont {Parkins},\ and\ \citenamefont
  {Dayan}}]{Rosenblum2011}%
  \BibitemOpen
  \bibfield  {author} {\bibinfo {author} {\bibfnamefont {S.}~\bibnamefont
  {Rosenblum}}, \bibinfo {author} {\bibfnamefont {S.}~\bibnamefont {Parkins}},
  \ and\ \bibinfo {author} {\bibfnamefont {B.}~\bibnamefont {Dayan}},\
  }\href@noop {} {\bibfield  {journal} {\bibinfo  {journal} {Phys. Rev. A},\
  }\textbf {\bibinfo {volume} {84}},\ \bibinfo {pages} {033854} (\bibinfo
  {year} {2011})}\BibitemShut {NoStop}%
\bibitem [{\citenamefont {O'Brien}\ \emph {et~al.}(2009)\citenamefont
  {O'Brien}, \citenamefont {Furusawa},\ and\ \citenamefont
  {Vu\v{c}kovi\'{c}}}]{Vuckovic2009}%
  \BibitemOpen
  \bibfield  {author} {\bibinfo {author} {\bibfnamefont {J.~L.}\ \bibnamefont
  {O'Brien}}, \bibinfo {author} {\bibfnamefont {A.}~\bibnamefont {Furusawa}}, \
  and\ \bibinfo {author} {\bibfnamefont {J.}~\bibnamefont {Vu\v{c}kovi\'{c}}},\
  }\href@noop {} {\bibfield  {journal} {\bibinfo  {journal} {Nat. Photonics},\
  }\textbf {\bibinfo {volume} {3}},\ \bibinfo {pages} {687} (\bibinfo {year}
  {2009})}\BibitemShut {NoStop}%
\bibitem [{\citenamefont {Tian}\ and\ \citenamefont
  {Carmichael}(1992)}]{Carmichael1992}%
  \BibitemOpen
  \bibfield  {author} {\bibinfo {author} {\bibfnamefont {L.}~\bibnamefont
  {Tian}}\ and\ \bibinfo {author} {\bibfnamefont {H.~J.}\ \bibnamefont
  {Carmichael}},\ }\href@noop {} {\bibfield  {journal} {\bibinfo  {journal}
  {Phys. Rev. A},\ }\textbf {\bibinfo {volume} {46}},\ \bibinfo {pages} {6801}
  (\bibinfo {year} {1992})}\BibitemShut {NoStop}%
\bibitem [{\citenamefont {Birnbaum}\ \emph {et~al.}(2005)\citenamefont
  {Birnbaum}, \citenamefont {Boca}, \citenamefont {Miller}, \citenamefont
  {Boozer}, \citenamefont {Northup},\ and\ \citenamefont
  {Kimble}}]{Kimble2005}%
  \BibitemOpen
  \bibfield  {author} {\bibinfo {author} {\bibfnamefont {K.~M.}\ \bibnamefont
  {Birnbaum}}, \bibinfo {author} {\bibfnamefont {A.}~\bibnamefont {Boca}},
  \bibinfo {author} {\bibfnamefont {R.}~\bibnamefont {Miller}}, \bibinfo
  {author} {\bibfnamefont {A.~D.}\ \bibnamefont {Boozer}}, \bibinfo {author}
  {\bibfnamefont {T.~E.}\ \bibnamefont {Northup}}, \ and\ \bibinfo {author}
  {\bibfnamefont {H.~J.}\ \bibnamefont {Kimble}},\ }\href@noop {} {\bibfield
  {journal} {\bibinfo  {journal} {Nature},\ }\textbf {\bibinfo {volume}
  {436}},\ \bibinfo {pages} {87} (\bibinfo {year} {2005})}\BibitemShut
  {NoStop}%
\bibitem [{\citenamefont {Dayan}\ \emph {et~al.}(2008)\citenamefont {Dayan},
  \citenamefont {Parkins}, \citenamefont {Aok}, \citenamefont {Ostby},
  \citenamefont {Vahala},\ and\ \citenamefont {Kimble}}]{Kimble2008}%
  \BibitemOpen
  \bibfield  {author} {\bibinfo {author} {\bibfnamefont {B.}~\bibnamefont
  {Dayan}}, \bibinfo {author} {\bibfnamefont {A.~S.}\ \bibnamefont {Parkins}},
  \bibinfo {author} {\bibfnamefont {T.}~\bibnamefont {Aok}}, \bibinfo {author}
  {\bibfnamefont {E.}~\bibnamefont {Ostby}}, \bibinfo {author} {\bibfnamefont
  {K.}~\bibnamefont {Vahala}}, \ and\ \bibinfo {author} {\bibfnamefont
  {H.}~\bibnamefont {Kimble}},\ }\href@noop {} {\bibfield  {journal} {\bibinfo
  {journal} {Science},\ }\textbf {\bibinfo {volume} {319}},\ \bibinfo {pages}
  {1062} (\bibinfo {year} {2008})}\BibitemShut {NoStop}%
\bibitem [{\citenamefont {Majumdar}\ \emph
  {et~al.}(2012){\natexlab{b}}\citenamefont {Majumdar}, \citenamefont {Bajcsy},
  \citenamefont {Rundquist},\ and\ \citenamefont
  {Vu\v{c}kovi\'{c}}}]{Majumdar2012}%
  \BibitemOpen
  \bibfield  {author} {\bibinfo {author} {\bibfnamefont {A.}~\bibnamefont
  {Majumdar}}, \bibinfo {author} {\bibfnamefont {M.}~\bibnamefont {Bajcsy}},
  \bibinfo {author} {\bibfnamefont {A.}~\bibnamefont {Rundquist}}, \ and\
  \bibinfo {author} {\bibfnamefont {J.}~\bibnamefont {Vu\v{c}kovi\'{c}}},\
  }\href@noop {} {\bibfield  {journal} {\bibinfo  {journal} {Phys. Rev.
  Lett.},\ }\textbf {\bibinfo {volume} {108}},\ \bibinfo {pages} {183601}
  (\bibinfo {year} {2012}{\natexlab{b}})}\BibitemShut {NoStop}%
\bibitem [{\citenamefont {Liew}\ and\ \citenamefont {Savona}(2010)}]{Liew2010}%
  \BibitemOpen
  \bibfield  {author} {\bibinfo {author} {\bibfnamefont {T.~C.~H.}\
  \bibnamefont {Liew}}\ and\ \bibinfo {author} {\bibfnamefont {V.}~\bibnamefont
  {Savona}},\ }\href@noop {} {\bibfield  {journal} {\bibinfo  {journal} {Phys.
  Rev. Lett.},\ }\textbf {\bibinfo {volume} {104}} (\bibinfo {year}
  {2010})}\BibitemShut {NoStop}%
\bibitem [{\citenamefont {Chang}\ \emph {et~al.}(2007)\citenamefont {Chang},
  \citenamefont {S{\o}rensen}, \citenamefont {Demler},\ and\ \citenamefont
  {Lukin}}]{Chang2007}%
  \BibitemOpen
  \bibfield  {author} {\bibinfo {author} {\bibfnamefont {D.~E.}\ \bibnamefont
  {Chang}}, \bibinfo {author} {\bibfnamefont {A.~S.}\ \bibnamefont
  {S{\o}rensen}}, \bibinfo {author} {\bibfnamefont {E.~A.}\ \bibnamefont
  {Demler}}, \ and\ \bibinfo {author} {\bibfnamefont {M.~D.}\ \bibnamefont
  {Lukin}},\ }\href@noop {} {\bibfield  {journal} {\bibinfo  {journal} {Nat.
  Phys.},\ }\textbf {\bibinfo {volume} {3}},\ \bibinfo {pages} {807} (\bibinfo
  {year} {2007})}\BibitemShut {NoStop}%
\bibitem [{\citenamefont {Gorshkov}\ \emph {et~al.}(2011)\citenamefont
  {Gorshkov}, \citenamefont {Otterbach}, \citenamefont {Fleischhauer},
  \citenamefont {Pohl},\ and\ \citenamefont {Lukin}}]{Gorshkov2011}%
  \BibitemOpen
  \bibfield  {author} {\bibinfo {author} {\bibfnamefont {A.~V.}\ \bibnamefont
  {Gorshkov}}, \bibinfo {author} {\bibfnamefont {J.}~\bibnamefont {Otterbach}},
  \bibinfo {author} {\bibfnamefont {M.}~\bibnamefont {Fleischhauer}}, \bibinfo
  {author} {\bibfnamefont {T.}~\bibnamefont {Pohl}}, \ and\ \bibinfo {author}
  {\bibfnamefont {M.~D.}\ \bibnamefont {Lukin}},\ }\href@noop {} {\bibfield
  {journal} {\bibinfo  {journal} {Phys. Rev. Lett.},\ }\textbf {\bibinfo
  {volume} {107}},\ \bibinfo {pages} {133602} (\bibinfo {year}
  {2011})}\BibitemShut {NoStop}%
\bibitem [{\citenamefont {Dudin}\ and\ \citenamefont
  {Kuzmich}(2012)}]{Dudin2012}%
  \BibitemOpen
  \bibfield  {author} {\bibinfo {author} {\bibfnamefont {Y.~O.}\ \bibnamefont
  {Dudin}}\ and\ \bibinfo {author} {\bibfnamefont {A.}~\bibnamefont
  {Kuzmich}},\ }\href@noop {} {\bibfield  {journal} {\bibinfo  {journal}
  {Science},\ }\textbf {\bibinfo {volume} {336}},\ \bibinfo {pages} {887}
  (\bibinfo {year} {2012})}\BibitemShut {NoStop}%
\bibitem [{\citenamefont {Peyronel}\ \emph {et~al.}(2012)\citenamefont
  {Peyronel}, \citenamefont {Firstenberg}, \citenamefont {Liang}, \citenamefont
  {Hofferberth}, \citenamefont {Gorshkov}, \citenamefont {Pohl}, \citenamefont
  {Lukin},\ and\ \citenamefont {Vuleti\'{c}}}]{Peyronel2012}%
  \BibitemOpen
  \bibfield  {author} {\bibinfo {author} {\bibfnamefont {T.}~\bibnamefont
  {Peyronel}}, \bibinfo {author} {\bibfnamefont {O.}~\bibnamefont
  {Firstenberg}}, \bibinfo {author} {\bibfnamefont {Q.}~\bibnamefont {Liang}},
  \bibinfo {author} {\bibfnamefont {S.}~\bibnamefont {Hofferberth}}, \bibinfo
  {author} {\bibfnamefont {A.~V.}\ \bibnamefont {Gorshkov}}, \bibinfo {author}
  {\bibfnamefont {T.}~\bibnamefont {Pohl}}, \bibinfo {author} {\bibfnamefont
  {M.~D.}\ \bibnamefont {Lukin}}, \ and\ \bibinfo {author} {\bibfnamefont
  {V.}~\bibnamefont {Vuleti\'{c}}},\ }\href@noop {} {\bibfield  {journal}
  {\bibinfo  {journal} {Nature},\ }\textbf {\bibinfo {volume} {488}},\ \bibinfo
  {pages} {57} (\bibinfo {year} {2012})}\BibitemShut {NoStop}%
\bibitem [{\citenamefont {Chang}\ \emph {et~al.}(2008)\citenamefont {Chang},
  \citenamefont {Gritsev}, \citenamefont {Morigi}, \citenamefont {Vuletic},
  \citenamefont {Lukin},\ and\ \citenamefont {Demler}}]{Chang2008}%
  \BibitemOpen
  \bibfield  {author} {\bibinfo {author} {\bibfnamefont {D.}~\bibnamefont
  {Chang}}, \bibinfo {author} {\bibfnamefont {V.}~\bibnamefont {Gritsev}},
  \bibinfo {author} {\bibfnamefont {G.}~\bibnamefont {Morigi}}, \bibinfo
  {author} {\bibfnamefont {V.}~\bibnamefont {Vuletic}}, \bibinfo {author}
  {\bibfnamefont {M.}~\bibnamefont {Lukin}}, \ and\ \bibinfo {author}
  {\bibfnamefont {E.}~\bibnamefont {Demler}},\ }\Doi {doi:10.1038/nphys1074}
  {\bibfield  {journal} {\bibinfo  {journal} {Nature Physics},\ }\textbf
  {\bibinfo {volume} {4}},\ \bibinfo {pages} {884 } (\bibinfo {year}
  {2008})}\BibitemShut {NoStop}%
\bibitem [{\citenamefont {Reinhard}\ \emph {et~al.}(2012)\citenamefont
  {Reinhard}, \citenamefont {Volz}, \citenamefont {Badolato}, \citenamefont
  {Hennessy}, \citenamefont {Hu},\ and\ \citenamefont
  {Imamo\u{g}lu}}]{Reinhard2012}%
  \BibitemOpen
  \bibfield  {author} {\bibinfo {author} {\bibfnamefont {A.}~\bibnamefont
  {Reinhard}}, \bibinfo {author} {\bibfnamefont {T.}~\bibnamefont {Volz}},
  \bibinfo {author} {\bibfnamefont {A.}~\bibnamefont {Badolato}}, \bibinfo
  {author} {\bibfnamefont {K.~J.}\ \bibnamefont {Hennessy}}, \bibinfo {author}
  {\bibfnamefont {E.~L.}\ \bibnamefont {Hu}}, \ and\ \bibinfo {author}
  {\bibfnamefont {A.}~\bibnamefont {Imamo\u{g}lu}},\ }\href@noop {} {\bibfield
  {journal} {\bibinfo  {journal} {Nat. Photonics},\ }\textbf {\bibinfo {volume}
  {6}},\ \bibinfo {pages} {93} (\bibinfo {year} {2012})}\BibitemShut {NoStop}%
\bibitem [{\citenamefont {Bamba}\ \emph {et~al.}(2011)\citenamefont {Bamba},
  \citenamefont {Imamo\u{g}lu}, \citenamefont {Carusotto},\ and\ \citenamefont
  {Ciuti}}]{Bamba2011}%
  \BibitemOpen
  \bibfield  {author} {\bibinfo {author} {\bibfnamefont {M.}~\bibnamefont
  {Bamba}}, \bibinfo {author} {\bibfnamefont {A.}~\bibnamefont {Imamo\u{g}lu}},
  \bibinfo {author} {\bibfnamefont {I.}~\bibnamefont {Carusotto}}, \ and\
  \bibinfo {author} {\bibfnamefont {C.}~\bibnamefont {Ciuti}},\ }\href@noop {}
  {\bibfield  {journal} {\bibinfo  {journal} {Phys. Rev. A},\ }\textbf
  {\bibinfo {volume} {83}},\ \bibinfo {pages} {021802(R)} (\bibinfo {year}
  {2011})}\BibitemShut {NoStop}%
\bibitem [{\citenamefont {Faraon}\ \emph
  {et~al.}(2011){\natexlab{a}}\citenamefont {Faraon}, \citenamefont {Majumdar},
  \citenamefont {Englund}, \citenamefont {Kim}, \citenamefont {Bajcsy},\ and\
  \citenamefont {Vu\v{c}kovi\'{c}}}]{Faraon2011}%
  \BibitemOpen
  \bibfield  {author} {\bibinfo {author} {\bibfnamefont {A.}~\bibnamefont
  {Faraon}}, \bibinfo {author} {\bibfnamefont {A.}~\bibnamefont {Majumdar}},
  \bibinfo {author} {\bibfnamefont {D.}~\bibnamefont {Englund}}, \bibinfo
  {author} {\bibfnamefont {E.}~\bibnamefont {Kim}}, \bibinfo {author}
  {\bibfnamefont {M.}~\bibnamefont {Bajcsy}}, \ and\ \bibinfo {author}
  {\bibfnamefont {J.}~\bibnamefont {Vu\v{c}kovi\'{c}}},\ }\href@noop {}
  {\bibfield  {journal} {\bibinfo  {journal} {New Journal of Physics},\
  }\textbf {\bibinfo {volume} {13}},\ \bibinfo {pages} {055025} (\bibinfo
  {year} {2011}{\natexlab{a}})}\BibitemShut {NoStop}%
\bibitem [{\citenamefont {Vu\v{c}kovi\'{c}}\ \emph {et~al.}(2001)\citenamefont
  {Vu\v{c}kovi\'{c}}, \citenamefont {Lon\v{c}ar}, \citenamefont {Mabuchi},\
  and\ \citenamefont {Scherer}}]{Vuckovic2001}%
  \BibitemOpen
  \bibfield  {author} {\bibinfo {author} {\bibfnamefont {J.}~\bibnamefont
  {Vu\v{c}kovi\'{c}}}, \bibinfo {author} {\bibfnamefont {M.}~\bibnamefont
  {Lon\v{c}ar}}, \bibinfo {author} {\bibfnamefont {H.}~\bibnamefont {Mabuchi}},
  \ and\ \bibinfo {author} {\bibfnamefont {A.}~\bibnamefont {Scherer}},\
  }\href@noop {} {\bibfield  {journal} {\bibinfo  {journal} {Phys. Rev. E},\
  }\textbf {\bibinfo {volume} {65}},\ \bibinfo {pages} {016608} (\bibinfo
  {year} {2001})}\BibitemShut {NoStop}%
\bibitem [{\citenamefont {Lev}\ \emph {et~al.}(2004)\citenamefont {Lev},
  \citenamefont {Srinivasan}, \citenamefont {Barclay}, \citenamefont
  {Painter},\ and\ \citenamefont {Mabuchi}}]{Lev2004}%
  \BibitemOpen
  \bibfield  {author} {\bibinfo {author} {\bibfnamefont {B.}~\bibnamefont
  {Lev}}, \bibinfo {author} {\bibfnamefont {K.}~\bibnamefont {Srinivasan}},
  \bibinfo {author} {\bibfnamefont {P.}~\bibnamefont {Barclay}}, \bibinfo
  {author} {\bibfnamefont {O.}~\bibnamefont {Painter}}, \ and\ \bibinfo
  {author} {\bibfnamefont {H.}~\bibnamefont {Mabuchi}},\ }\href@noop {}
  {\bibfield  {journal} {\bibinfo  {journal} {Nanotechnology},\ }\textbf
  {\bibinfo {volume} {15}},\ \bibinfo {pages} {S556} (\bibinfo {year}
  {2004})}\BibitemShut {NoStop}%
\bibitem [{\citenamefont {Yamamoto}\ \emph {et~al.}(2008)\citenamefont
  {Yamamoto}, \citenamefont {Notomi}, \citenamefont {Taniyama}, \citenamefont
  {Kuramochi}, \citenamefont {Yoshikawa}, \citenamefont {Torii},\ and\
  \citenamefont {Kuga}}]{Yamamoto2008}%
  \BibitemOpen
  \bibfield  {author} {\bibinfo {author} {\bibfnamefont {T.}~\bibnamefont
  {Yamamoto}}, \bibinfo {author} {\bibfnamefont {M.}~\bibnamefont {Notomi}},
  \bibinfo {author} {\bibfnamefont {H.}~\bibnamefont {Taniyama}}, \bibinfo
  {author} {\bibfnamefont {E.}~\bibnamefont {Kuramochi}}, \bibinfo {author}
  {\bibfnamefont {Y.}~\bibnamefont {Yoshikawa}}, \bibinfo {author}
  {\bibfnamefont {Y.}~\bibnamefont {Torii}}, \ and\ \bibinfo {author}
  {\bibfnamefont {T.}~\bibnamefont {Kuga}},\ }\href@noop {} {\bibfield
  {journal} {\bibinfo  {journal} {Optics Express},\ }\textbf {\bibinfo {volume}
  {16}},\ \bibinfo {pages} {13809} (\bibinfo {year} {2008})}\BibitemShut
  {NoStop}%
\bibitem [{\citenamefont {Quan}\ and\ \citenamefont {Loncar}(2011)}]{Quan2011}%
  \BibitemOpen
  \bibfield  {author} {\bibinfo {author} {\bibfnamefont {Q.}~\bibnamefont
  {Quan}}\ and\ \bibinfo {author} {\bibfnamefont {M.}~\bibnamefont {Loncar}},\
  }\href@noop {} {\bibfield  {journal} {\bibinfo  {journal} {Optics Express},\
  }\textbf {\bibinfo {volume} {19}},\ \bibinfo {pages} {18529} (\bibinfo {year}
  {2011})}\BibitemShut {NoStop}%
\bibitem [{\citenamefont {Li}\ \emph {et~al.}(2011)\citenamefont {Li},
  \citenamefont {Fattal}, \citenamefont {Fiorentino},\ and\ \citenamefont
  {Beausoleil}}]{Li2011}%
  \BibitemOpen
  \bibfield  {author} {\bibinfo {author} {\bibfnamefont {J.}~\bibnamefont
  {Li}}, \bibinfo {author} {\bibfnamefont {D.}~\bibnamefont {Fattal}}, \bibinfo
  {author} {\bibfnamefont {M.}~\bibnamefont {Fiorentino}}, \ and\ \bibinfo
  {author} {\bibfnamefont {R.~G.}\ \bibnamefont {Beausoleil}},\ }\Doi
  {10.1103/PhysRevLett.106.193901} {\bibfield  {journal} {\bibinfo  {journal}
  {Phys. Rev. Lett.},\ }\textbf {\bibinfo {volume} {106}},\ \bibinfo {pages}
  {193901} (\bibinfo {year} {2011})}\BibitemShut {NoStop}%
\bibitem [{\citenamefont {Werner}\ and\ \citenamefont
  {Imamo\u{g}lu}(1999)}]{Werner1999}%
  \BibitemOpen
  \bibfield  {author} {\bibinfo {author} {\bibfnamefont {M.~J.}\ \bibnamefont
  {Werner}}\ and\ \bibinfo {author} {\bibfnamefont {A.}~\bibnamefont
  {Imamo\u{g}lu}},\ }\href@noop {} {\bibfield  {journal} {\bibinfo  {journal}
  {Phys. Rev. A},\ }\textbf {\bibinfo {volume} {61}},\ \bibinfo {pages}
  {011801(R)} (\bibinfo {year} {1999})}\BibitemShut {NoStop}%
\bibitem [{\citenamefont {Rebi\'{c}}\ \emph {et~al.}(1999)\citenamefont
  {Rebi\'{c}}, \citenamefont {Tan}, \citenamefont {Parkins},\ and\
  \citenamefont {Walls}}]{Rebic1999}%
  \BibitemOpen
  \bibfield  {author} {\bibinfo {author} {\bibfnamefont {S.}~\bibnamefont
  {Rebi\'{c}}}, \bibinfo {author} {\bibfnamefont {S.~M.}\ \bibnamefont {Tan}},
  \bibinfo {author} {\bibfnamefont {A.}~\bibnamefont {Parkins}}, \ and\
  \bibinfo {author} {\bibfnamefont {D.~F.}\ \bibnamefont {Walls}},\ }\href@noop
  {} {\bibfield  {journal} {\bibinfo  {journal} {J. Opt. B: Quantum Semiclass.
  Opt},\ \bibinfo {pages} {490}} (\bibinfo {year} {1999})}\BibitemShut
  {NoStop}%
\bibitem [{\citenamefont {Gheri}\ \emph {et~al.}(1999)\citenamefont {Gheri},
  \citenamefont {Alge},\ and\ \citenamefont {Grangier}}]{Gheri1999}%
  \BibitemOpen
  \bibfield  {author} {\bibinfo {author} {\bibfnamefont {K.~M.}\ \bibnamefont
  {Gheri}}, \bibinfo {author} {\bibfnamefont {W.}~\bibnamefont {Alge}}, \ and\
  \bibinfo {author} {\bibfnamefont {P.}~\bibnamefont {Grangier}},\ }\href@noop
  {} {\bibfield  {journal} {\bibinfo  {journal} {Phys. Rev. A},\ }\textbf
  {\bibinfo {volume} {60}},\ \bibinfo {pages} {R2673} (\bibinfo {year}
  {1999})}\BibitemShut {NoStop}%
\bibitem [{\citenamefont {Greentree}\ \emph {et~al.}(2000)\citenamefont
  {Greentree}, \citenamefont {Vaccaro}, \citenamefont {Echaniz}, \citenamefont
  {Durrant},\ and\ \citenamefont {Marangos}}]{Greentree2000}%
  \BibitemOpen
  \bibfield  {author} {\bibinfo {author} {\bibfnamefont {A.~D.}\ \bibnamefont
  {Greentree}}, \bibinfo {author} {\bibfnamefont {J.~A.}\ \bibnamefont
  {Vaccaro}}, \bibinfo {author} {\bibfnamefont {R.~D.}\ \bibnamefont
  {Echaniz}}, \bibinfo {author} {\bibfnamefont {A.~V.}\ \bibnamefont
  {Durrant}}, \ and\ \bibinfo {author} {\bibfnamefont {J.~P.}\ \bibnamefont
  {Marangos}},\ }\href@noop {} {\bibfield  {journal} {\bibinfo  {journal} {J.
  Opt. B: Quantum Semiclass. Opt.},\ }\textbf {\bibinfo {volume} {2}},\
  \bibinfo {pages} {252} (\bibinfo {year} {2000})}\BibitemShut {NoStop}%
\bibitem [{\citenamefont {Rebi\'{c}}\ \emph
  {et~al.}(2002){\natexlab{a}}\citenamefont {Rebi\'{c}}, \citenamefont
  {Parkins},\ and\ \citenamefont {Tan}}]{Rebic2002a}%
  \BibitemOpen
  \bibfield  {author} {\bibinfo {author} {\bibfnamefont {S.}~\bibnamefont
  {Rebi\'{c}}}, \bibinfo {author} {\bibfnamefont {A.~S.}\ \bibnamefont
  {Parkins}}, \ and\ \bibinfo {author} {\bibfnamefont {S.~M.}\ \bibnamefont
  {Tan}},\ }\href@noop {} {\bibfield  {journal} {\bibinfo  {journal} {Phys.
  Rev. A},\ }\textbf {\bibinfo {volume} {65}},\ \bibinfo {pages} {043806}
  (\bibinfo {year} {2002}{\natexlab{a}})}\BibitemShut {NoStop}%
\bibitem [{\citenamefont {Rebi\'{c}}\ \emph
  {et~al.}(2002){\natexlab{b}}\citenamefont {Rebi\'{c}}, \citenamefont
  {Parkins},\ and\ \citenamefont {Tan}}]{Rebic2002b}%
  \BibitemOpen
  \bibfield  {author} {\bibinfo {author} {\bibfnamefont {S.}~\bibnamefont
  {Rebi\'{c}}}, \bibinfo {author} {\bibfnamefont {A.~S.}\ \bibnamefont
  {Parkins}}, \ and\ \bibinfo {author} {\bibfnamefont {S.~M.}\ \bibnamefont
  {Tan}},\ }\href@noop {} {\bibfield  {journal} {\bibinfo  {journal} {Phys.
  Rev. A},\ }\textbf {\bibinfo {volume} {65}},\ \bibinfo {pages} {063804}
  (\bibinfo {year} {2002}{\natexlab{b}})}\BibitemShut {NoStop}%
\bibitem [{\citenamefont {Hennessy}\ \emph {et~al.}(2007)\citenamefont
  {Hennessy}, \citenamefont {Badolato}, \citenamefont {Winger}, \citenamefont
  {Gerace}, \citenamefont {Atat\"{u}re}, \citenamefont {Gulde}, \citenamefont
  {F\"{a}lt}, \citenamefont {Hu},\ and\ \citenamefont
  {Imamo\u{g}lu}}]{Hennessy2007}%
  \BibitemOpen
  \bibfield  {author} {\bibinfo {author} {\bibfnamefont {K.}~\bibnamefont
  {Hennessy}}, \bibinfo {author} {\bibfnamefont {A.}~\bibnamefont {Badolato}},
  \bibinfo {author} {\bibfnamefont {M.}~\bibnamefont {Winger}}, \bibinfo
  {author} {\bibfnamefont {D.}~\bibnamefont {Gerace}}, \bibinfo {author}
  {\bibfnamefont {M.}~\bibnamefont {Atat\"{u}re}}, \bibinfo {author}
  {\bibfnamefont {S.}~\bibnamefont {Gulde}}, \bibinfo {author} {\bibfnamefont
  {S.}~\bibnamefont {F\"{a}lt}}, \bibinfo {author} {\bibfnamefont {E.~L.}\
  \bibnamefont {Hu}}, \ and\ \bibinfo {author} {\bibfnamefont {A.}~\bibnamefont
  {Imamo\u{g}lu}},\ }\href@noop {} {\bibfield  {journal} {\bibinfo  {journal}
  {Nature},\ }\textbf {\bibinfo {volume} {445}},\ \bibinfo {pages} {896}
  (\bibinfo {year} {2007})}\BibitemShut {NoStop}%
\bibitem [{\citenamefont {Dawkins}\ \emph {et~al.}(2011)\citenamefont
  {Dawkins}, \citenamefont {Mitsch}, \citenamefont {Reitz}, \citenamefont
  {Vetsch},\ and\ \citenamefont {Rauschenbeutel}}]{Dawkins2011}%
  \BibitemOpen
  \bibfield  {author} {\bibinfo {author} {\bibfnamefont {S.~T.}\ \bibnamefont
  {Dawkins}}, \bibinfo {author} {\bibfnamefont {R.}~\bibnamefont {Mitsch}},
  \bibinfo {author} {\bibfnamefont {D.}~\bibnamefont {Reitz}}, \bibinfo
  {author} {\bibfnamefont {E.}~\bibnamefont {Vetsch}}, \ and\ \bibinfo {author}
  {\bibfnamefont {A.}~\bibnamefont {Rauschenbeutel}},\ }\href@noop {}
  {\bibfield  {journal} {\bibinfo  {journal} {Phys. Rev. Lett.},\ }\textbf
  {\bibinfo {volume} {107}},\ \bibinfo {pages} {243601} (\bibinfo {year}
  {2011})}\BibitemShut {NoStop}%
\bibitem [{\citenamefont {Lacro\^{u}te}\ \emph {et~al.}(2012)\citenamefont
  {Lacro\^{u}te}, \citenamefont {Choi}, \citenamefont {Goban}, \citenamefont
  {Alton}, \citenamefont {Ding}, \citenamefont {Stern},\ and\ \citenamefont
  {Kimble}}]{Lacroute2012}%
  \BibitemOpen
  \bibfield  {author} {\bibinfo {author} {\bibfnamefont {C.}~\bibnamefont
  {Lacro\^{u}te}}, \bibinfo {author} {\bibfnamefont {K.~S.}\ \bibnamefont
  {Choi}}, \bibinfo {author} {\bibfnamefont {A.}~\bibnamefont {Goban}},
  \bibinfo {author} {\bibfnamefont {D.~J.}\ \bibnamefont {Alton}}, \bibinfo
  {author} {\bibfnamefont {D.}~\bibnamefont {Ding}}, \bibinfo {author}
  {\bibfnamefont {N.~P.}\ \bibnamefont {Stern}}, \ and\ \bibinfo {author}
  {\bibfnamefont {H.~J.}\ \bibnamefont {Kimble}},\ }\href@noop {} {\bibfield
  {journal} {\bibinfo  {journal} {New Journal of Physics},\ }\textbf {\bibinfo
  {volume} {14}},\ \bibinfo {pages} {023056} (\bibinfo {year}
  {2012})}\BibitemShut {NoStop}%
\bibitem [{\citenamefont {Tan}(1999)}]{qotoolbox}%
  \BibitemOpen
  \bibfield  {author} {\bibinfo {author} {\bibfnamefont {S.~M.}\ \bibnamefont
  {Tan}},\ }\href@noop {} {\bibfield  {journal} {\bibinfo  {journal} {J. Opt.
  B: Quantum Semiclass. Opt},\ }\textbf {\bibinfo {volume} {1}},\ \bibinfo
  {pages} {424} (\bibinfo {year} {1999})}\BibitemShut {NoStop}%
\bibitem [{\citenamefont {Hood}\ \emph {et~al.}(1998)\citenamefont {Hood},
  \citenamefont {Chapman}, \citenamefont {Lynn},\ and\ \citenamefont
  {Kimble}}]{Hood1998}%
  \BibitemOpen
  \bibfield  {author} {\bibinfo {author} {\bibfnamefont {C.~J.}\ \bibnamefont
  {Hood}}, \bibinfo {author} {\bibfnamefont {M.~S.}\ \bibnamefont {Chapman}},
  \bibinfo {author} {\bibfnamefont {T.~W.}\ \bibnamefont {Lynn}}, \ and\
  \bibinfo {author} {\bibfnamefont {H.~J.}\ \bibnamefont {Kimble}},\
  }\href@noop {} {\bibfield  {journal} {\bibinfo  {journal} {Phys. Rev.
  Lett.},\ }\textbf {\bibinfo {volume} {80}},\ \bibinfo {pages} {4157}
  (\bibinfo {year} {1998})}\BibitemShut {NoStop}%
\bibitem [{\citenamefont {Rivoire}\ \emph {et~al.}(2008)\citenamefont
  {Rivoire}, \citenamefont {Faraon},\ and\ \citenamefont
  {Vu\v{c}kovi\'{c}}}]{Rivoire2008}%
  \BibitemOpen
  \bibfield  {author} {\bibinfo {author} {\bibfnamefont {K.}~\bibnamefont
  {Rivoire}}, \bibinfo {author} {\bibfnamefont {A.}~\bibnamefont {Faraon}}, \
  and\ \bibinfo {author} {\bibfnamefont {J.}~\bibnamefont {Vu\v{c}kovi\'{c}}},\
  }\href@noop {} {\bibfield  {journal} {\bibinfo  {journal} {Appl. Phys.
  Lett.},\ }\textbf {\bibinfo {volume} {93}},\ \bibinfo {pages} {063103}
  (\bibinfo {year} {2008})}\BibitemShut {NoStop}%
\bibitem [{\citenamefont {Gong}\ \emph {et~al.}(2010)\citenamefont {Gong},
  \citenamefont {Ishikawa}, \citenamefont {Cheng}, \citenamefont {Gunji},
  \citenamefont {Nishi},\ and\ \citenamefont {Vu\v{c}kovi\'{c}}}]{Gong2010}%
  \BibitemOpen
  \bibfield  {author} {\bibinfo {author} {\bibfnamefont {Y.}~\bibnamefont
  {Gong}}, \bibinfo {author} {\bibfnamefont {S.}~\bibnamefont {Ishikawa}},
  \bibinfo {author} {\bibfnamefont {S.-L.}\ \bibnamefont {Cheng}}, \bibinfo
  {author} {\bibfnamefont {M.}~\bibnamefont {Gunji}}, \bibinfo {author}
  {\bibfnamefont {Y.}~\bibnamefont {Nishi}}, \ and\ \bibinfo {author}
  {\bibfnamefont {J.}~\bibnamefont {Vu\v{c}kovi\'{c}}},\ }\href@noop {}
  {\bibfield  {journal} {\bibinfo  {journal} {Phys. Rev. B},\ }\textbf
  {\bibinfo {volume} {81}},\ \bibinfo {pages} {235317} (\bibinfo {year}
  {2010})}\BibitemShut {NoStop}%
\bibitem [{\citenamefont {Englund}\ \emph {et~al.}(2007)\citenamefont
  {Englund}, \citenamefont {Faraon}, \citenamefont {Fushman}, \citenamefont
  {Stoltz}, \citenamefont {Petroff},\ and\ \citenamefont
  {Vu\v{c}kovi\'{c}}}]{Englund2007}%
  \BibitemOpen
  \bibfield  {author} {\bibinfo {author} {\bibfnamefont {D.}~\bibnamefont
  {Englund}}, \bibinfo {author} {\bibfnamefont {A.}~\bibnamefont {Faraon}},
  \bibinfo {author} {\bibfnamefont {I.}~\bibnamefont {Fushman}}, \bibinfo
  {author} {\bibfnamefont {N.}~\bibnamefont {Stoltz}}, \bibinfo {author}
  {\bibfnamefont {P.}~\bibnamefont {Petroff}}, \ and\ \bibinfo {author}
  {\bibfnamefont {J.}~\bibnamefont {Vu\v{c}kovi\'{c}}},\ }\href@noop {}
  {\bibfield  {journal} {\bibinfo  {journal} {Nature},\ }\textbf {\bibinfo
  {volume} {450}},\ \bibinfo {pages} {857} (\bibinfo {year}
  {2007})}\BibitemShut {NoStop}%
\bibitem [{\citenamefont {Waks}\ and\ \citenamefont
  {Vu\v{c}kovi\'{c}}(2006)}]{Waks2006}%
  \BibitemOpen
  \bibfield  {author} {\bibinfo {author} {\bibfnamefont {E.}~\bibnamefont
  {Waks}}\ and\ \bibinfo {author} {\bibfnamefont {J.}~\bibnamefont
  {Vu\v{c}kovi\'{c}}},\ }\href@noop {} {\bibfield  {journal} {\bibinfo
  {journal} {Phys. Rev. Lett.},\ }\textbf {\bibinfo {volume} {96}},\ \bibinfo
  {pages} {153601} (\bibinfo {year} {2006})}\BibitemShut {NoStop}%
\bibitem [{\citenamefont {Koch}\ \emph {et~al.}(2011)\citenamefont {Koch},
  \citenamefont {Sames}, \citenamefont {Balbach}, \citenamefont {Chibani},
  \citenamefont {Kubanek}, \citenamefont {Murr}, \citenamefont {Wilk},\ and\
  \citenamefont {Rempe}}]{Koch2011}%
  \BibitemOpen
  \bibfield  {author} {\bibinfo {author} {\bibfnamefont {M.}~\bibnamefont
  {Koch}}, \bibinfo {author} {\bibfnamefont {C.}~\bibnamefont {Sames}},
  \bibinfo {author} {\bibfnamefont {M.}~\bibnamefont {Balbach}}, \bibinfo
  {author} {\bibfnamefont {H.}~\bibnamefont {Chibani}}, \bibinfo {author}
  {\bibfnamefont {A.}~\bibnamefont {Kubanek}}, \bibinfo {author} {\bibfnamefont
  {K.}~\bibnamefont {Murr}}, \bibinfo {author} {\bibfnamefont {T.}~\bibnamefont
  {Wilk}}, \ and\ \bibinfo {author} {\bibfnamefont {G.}~\bibnamefont {Rempe}},\
  }\href@noop {} {\bibfield  {journal} {\bibinfo  {journal} {Phys. Rev.
  Lett.},\ }\textbf {\bibinfo {volume} {107}},\ \bibinfo {pages} {023601}
  (\bibinfo {year} {2011})}\BibitemShut {NoStop}%
\bibitem [{\citenamefont {Barclay}\ \emph {et~al.}(2006)\citenamefont
  {Barclay}, \citenamefont {Srinivasan}, \citenamefont {Painter}, \citenamefont
  {Lev},\ and\ \citenamefont {Mabuchi}}]{Barclay2006}%
  \BibitemOpen
  \bibfield  {author} {\bibinfo {author} {\bibfnamefont {P.~E.}\ \bibnamefont
  {Barclay}}, \bibinfo {author} {\bibfnamefont {K.}~\bibnamefont {Srinivasan}},
  \bibinfo {author} {\bibfnamefont {O.}~\bibnamefont {Painter}}, \bibinfo
  {author} {\bibfnamefont {B.}~\bibnamefont {Lev}}, \ and\ \bibinfo {author}
  {\bibfnamefont {H.}~\bibnamefont {Mabuchi}},\ }\href@noop {} {\bibfield
  {journal} {\bibinfo  {journal} {Appl. Phys. Lett.},\ }\textbf {\bibinfo
  {volume} {89}},\ \bibinfo {pages} {131108} (\bibinfo {year}
  {2006})}\BibitemShut {NoStop}%
\bibitem [{\citenamefont {Birnbaum}\ \emph {et~al.}(2006)\citenamefont
  {Birnbaum}, \citenamefont {Parkins},\ and\ \citenamefont
  {Kimble}}]{Birnbaum2006}%
  \BibitemOpen
  \bibfield  {author} {\bibinfo {author} {\bibfnamefont {K.~M.}\ \bibnamefont
  {Birnbaum}}, \bibinfo {author} {\bibfnamefont {A.~S.}\ \bibnamefont
  {Parkins}}, \ and\ \bibinfo {author} {\bibfnamefont {H.~J.}\ \bibnamefont
  {Kimble}},\ }\href@noop {} {\bibfield  {journal} {\bibinfo  {journal} {Phys.
  Rev. A},\ }\textbf {\bibinfo {volume} {74}},\ \bibinfo {pages} {063802}
  (\bibinfo {year} {2006})}\BibitemShut {NoStop}%
\bibitem [{\citenamefont {Rice}\ and\ \citenamefont
  {Carmichael}(1988)}]{Rice1988}%
  \BibitemOpen
  \bibfield  {author} {\bibinfo {author} {\bibfnamefont {P.~R.}\ \bibnamefont
  {Rice}}\ and\ \bibinfo {author} {\bibfnamefont {H.~J.}\ \bibnamefont
  {Carmichael}},\ }\href@noop {} {\bibfield  {journal} {\bibinfo  {journal}
  {IEEE Journal of Quantum Electronics},\ }\textbf {\bibinfo {volume} {24}},\
  \bibinfo {pages} {1351} (\bibinfo {year} {1988})}\BibitemShut {NoStop}%
\bibitem [{\citenamefont {Aoki}\ \emph {et~al.}(2009)\citenamefont {Aoki},
  \citenamefont {Parkins}, \citenamefont {Alton}, \citenamefont {Regal},
  \citenamefont {Dayan}, \citenamefont {Ostby}, \citenamefont {Vahala},\ and\
  \citenamefont {Kimble}}]{Aoki2009}%
  \BibitemOpen
  \bibfield  {author} {\bibinfo {author} {\bibfnamefont {T.}~\bibnamefont
  {Aoki}}, \bibinfo {author} {\bibfnamefont {A.~S.}\ \bibnamefont {Parkins}},
  \bibinfo {author} {\bibfnamefont {D.~J.}\ \bibnamefont {Alton}}, \bibinfo
  {author} {\bibfnamefont {C.~A.}\ \bibnamefont {Regal}}, \bibinfo {author}
  {\bibfnamefont {B.}~\bibnamefont {Dayan}}, \bibinfo {author} {\bibfnamefont
  {E.}~\bibnamefont {Ostby}}, \bibinfo {author} {\bibfnamefont {K.~J.}\
  \bibnamefont {Vahala}}, \ and\ \bibinfo {author} {\bibfnamefont {H.~J.}\
  \bibnamefont {Kimble}},\ }\href@noop {} {\bibfield  {journal} {\bibinfo
  {journal} {Phys. Rev. Lett.},\ }\textbf {\bibinfo {volume} {102}},\ \bibinfo
  {pages} {083601} (\bibinfo {year} {2009})}\BibitemShut {NoStop}%
\bibitem [{\citenamefont {Alton}\ \emph {et~al.}(2010)\citenamefont {Alton},
  \citenamefont {Stern}, \citenamefont {Aoki}, \citenamefont {Lee},
  \citenamefont {Ostby}, \citenamefont {Vahala},\ and\ \citenamefont
  {Kimble}}]{Alton2010}%
  \BibitemOpen
  \bibfield  {author} {\bibinfo {author} {\bibfnamefont {D.~J.}\ \bibnamefont
  {Alton}}, \bibinfo {author} {\bibfnamefont {N.~P.}\ \bibnamefont {Stern}},
  \bibinfo {author} {\bibfnamefont {T.}~\bibnamefont {Aoki}}, \bibinfo {author}
  {\bibfnamefont {H.}~\bibnamefont {Lee}}, \bibinfo {author} {\bibfnamefont
  {E.}~\bibnamefont {Ostby}}, \bibinfo {author} {\bibfnamefont {K.~J.}\
  \bibnamefont {Vahala}}, \ and\ \bibinfo {author} {\bibfnamefont {H.~J.}\
  \bibnamefont {Kimble}},\ }\href@noop {} {\bibfield  {journal} {\bibinfo
  {journal} {Nat. Phys.},\ }\textbf {\bibinfo {volume} {7}},\ \bibinfo {pages}
  {159} (\bibinfo {year} {2010})}\BibitemShut {NoStop}%
\bibitem [{\citenamefont {Faraon}\ \emph
  {et~al.}(2011){\natexlab{b}}\citenamefont {Faraon}, \citenamefont {Barclay},
  \citenamefont {Santori}, \citenamefont {Fu},\ and\ \citenamefont
  {Beausoleil}}]{Faraon2011a}%
  \BibitemOpen
  \bibfield  {author} {\bibinfo {author} {\bibfnamefont {A.}~\bibnamefont
  {Faraon}}, \bibinfo {author} {\bibfnamefont {P.~E.}\ \bibnamefont {Barclay}},
  \bibinfo {author} {\bibfnamefont {C.}~\bibnamefont {Santori}}, \bibinfo
  {author} {\bibfnamefont {K.-M.~C.}\ \bibnamefont {Fu}}, \ and\ \bibinfo
  {author} {\bibfnamefont {R.~G.}\ \bibnamefont {Beausoleil}},\ }\href@noop {}
  {\bibfield  {journal} {\bibinfo  {journal} {Nat. Photonics},\ }\textbf
  {\bibinfo {volume} {5}},\ \bibinfo {pages} {301} (\bibinfo {year}
  {2011}{\natexlab{b}})}\BibitemShut {NoStop}%
\bibitem [{\citenamefont {Babinec}\ \emph {et~al.}(2011)\citenamefont
  {Babinec}, \citenamefont {Choy}, \citenamefont {Smith}, \citenamefont
  {Khan},\ and\ \citenamefont {Lon\v{c}ar}}]{Babinec2011}%
  \BibitemOpen
  \bibfield  {author} {\bibinfo {author} {\bibfnamefont {T.~M.}\ \bibnamefont
  {Babinec}}, \bibinfo {author} {\bibfnamefont {J.~T.}\ \bibnamefont {Choy}},
  \bibinfo {author} {\bibfnamefont {K.~J.~M.}\ \bibnamefont {Smith}}, \bibinfo
  {author} {\bibfnamefont {M.}~\bibnamefont {Khan}}, \ and\ \bibinfo {author}
  {\bibfnamefont {M.}~\bibnamefont {Lon\v{c}ar}},\ }\href@noop {} {\bibfield
  {journal} {\bibinfo  {journal} {J. Vac. Sci. Technol. B},\ }\textbf {\bibinfo
  {volume} {29}},\ \bibinfo {pages} {010601} (\bibinfo {year}
  {2011})}\BibitemShut {NoStop}%
\bibitem [{\citenamefont {Englund}\ \emph {et~al.}(2010)\citenamefont
  {Englund}, \citenamefont {Shields}, \citenamefont {Rivoire}, \citenamefont
  {Hatami}, \citenamefont {Vu\v{c}kovi\'{c}}, \citenamefont {Park},\ and\
  \citenamefont {Lukin}}]{Englund2010}%
  \BibitemOpen
  \bibfield  {author} {\bibinfo {author} {\bibfnamefont {D.}~\bibnamefont
  {Englund}}, \bibinfo {author} {\bibfnamefont {B.}~\bibnamefont {Shields}},
  \bibinfo {author} {\bibfnamefont {K.}~\bibnamefont {Rivoire}}, \bibinfo
  {author} {\bibfnamefont {F.}~\bibnamefont {Hatami}}, \bibinfo {author}
  {\bibfnamefont {J.}~\bibnamefont {Vu\v{c}kovi\'{c}}}, \bibinfo {author}
  {\bibfnamefont {H.}~\bibnamefont {Park}}, \ and\ \bibinfo {author}
  {\bibfnamefont {M.}~\bibnamefont {Lukin}},\ }\href@noop {} {\bibfield
  {journal} {\bibinfo  {journal} {Nano Letters},\ }\textbf {\bibinfo {volume}
  {10}},\ \bibinfo {pages} {3922} (\bibinfo {year} {2010})}\BibitemShut
  {NoStop}%
\bibitem [{\citenamefont {Press}\ \emph {et~al.}(2008)\citenamefont {Press},
  \citenamefont {Ladd}, \citenamefont {Zhang},\ and\ \citenamefont
  {Yamamoto}}]{Press2008}%
  \BibitemOpen
  \bibfield  {author} {\bibinfo {author} {\bibfnamefont {D.}~\bibnamefont
  {Press}}, \bibinfo {author} {\bibfnamefont {T.~D.}\ \bibnamefont {Ladd}},
  \bibinfo {author} {\bibfnamefont {B.}~\bibnamefont {Zhang}}, \ and\ \bibinfo
  {author} {\bibfnamefont {Y.}~\bibnamefont {Yamamoto}},\ }\href@noop {}
  {\bibfield  {journal} {\bibinfo  {journal} {Nature},\ }\textbf {\bibinfo
  {volume} {456}},\ \bibinfo {pages} {218} (\bibinfo {year}
  {2008})}\BibitemShut {NoStop}%
\end{thebibliography}%

\end{document}